\documentclass[12pt,a4]{article}
\pdfoutput=1
\usepackage{setspace}
\usepackage{amsmath}
\usepackage{amssymb}
\usepackage{colortbl}
\usepackage{gensymb}
\usepackage[table,x11names]{xcolor}
\usepackage{graphicx}
\usepackage{hyperref}
\usepackage{cite}
\usepackage{slashed}
\usepackage{pifont}
\usepackage{longtable}
\usepackage{multirow}
\usepackage{ltablex}
\usepackage{array}
\usepackage{rotating}
\usepackage{booktabs}
\usepackage{hhline}
\usepackage{BOONDOX-cal}
\usepackage{bbm}
\definecolor{nicered}{rgb}{0.6,0,0}
\definecolor{nicegreen}{rgb}{0.1,0.5,0.1}
\definecolor{niceblue}{rgb}{0,0.4,0.8}
\hypersetup{colorlinks,citecolor= nicegreen,linkcolor=black,urlcolor=niceblue}

\allowdisplaybreaks

\begin{document}
\begin{titlepage}
% -------------------------------------------------
% Uncomment to add preprint numbers when necessary
% -------------------------------------------------

  \newcommand{\AddrLiege}{{\sl \small IFPA, D\'ep. AGO, Universit\'e de
      Li\`ege, B\^at B5, Sart Tilman B-4000 Li\`ege 1,
      Belgium}}
  \newcommand{\AddrUFSM}{{\sl \small  Universidad T\'ecnica 
      Federico Santa Mar\'{i}a - Departamento de F\'{i}sica\\
      Casilla 110-V, Avda. Espa\~na 1680, Valpara\'{i}so, Chile}}
  \newcommand{\AddrIFIC}{{\sl \small AHEP Group, Instituto de F\'{\i}sica Corpuscular,
CSIC/Universitat de Val\`encia,  \\
Calle Catedr\'atico Jos\'e Beltr\'an, 2 E-46980 Paterna, Spain}}
  \vspace*{0.5cm}
\begin{center}
  \textbf{\Large COHERENT analysis of\\[3.5mm]
    neutrino generalized interactions}
  \\[9mm]
  D. Aristizabal Sierra$^{a,b,}$\footnote{email address: {\tt
      daristizabal@ulg.ac.be}}, Valentina De
  Romeri$^{c,}$\footnote{email address: {\tt deromeri@ific.uv.es}},
  N. Rojas$^{a,}$\footnote{email address: {\tt
      nicolas.rojasro@usm.cl}}
  \vspace{0.8cm}\\
  $^a$\AddrUFSM\\[3mm]
  $^b$\AddrLiege\\[3mm]
  $^c$\AddrIFIC
\end{center}
\vspace*{0.2cm}
\begin{abstract}
  \onehalfspacing
  Effective neutrino-quark generalized interactions are entirely
  determined by Lorentz invariance, so they include all possible
  four-fermion non-derivative Lorentz structures. They contain
  neutrino-quark non-standard interactions as a subset, but span over
  a larger set that involves effective scalar, pseudoscalar, axial and
  tensor operators. Using recent COHERENT data, we derive constraints
  on the corresponding couplings by considering scalar, vector and
  tensor quark currents and assuming no lepton flavor dependence. We
  allow for mixed neutrino-quark Lorentz couplings and consider two
  types of scenarios in which: (i) one interaction at the nuclear level
  is present at a time, (ii) two interactions are simultaneously
  present. For scenarios (i) our findings show that scalar
  interactions are the most severely constrained, in particular for
  pseudoscalar-scalar neutrino-quark couplings. In contrast, tensor
  and non-standard vector interactions still enable for sizable
  effective parameters.  We find as well that an extra vector
  interaction improves the data fit when compared with the result
  derived assuming only the standard model contribution. In scenarios
  (ii) the presence of two interactions relaxes the bounds and opens
  regions in parameter space that are otherwise closed, with the
  effect being more pronounced in the scalar-vector and scalar-tensor
  cases. We point out that barring the vector case, our results
  represent the most stringent bounds on effective neutrino-quark
  generalized interactions for mediator masses of order $\sim 1$\,GeV.
  They hold as well for larger mediator masses, case in which they
  should be compared with limits from neutrino deep-inelastic
  scattering data.
\end{abstract}
\end{titlepage}
\setcounter{footnote}{0}
\tableofcontents
\vspace{1cm}
\section{Introduction}
\label{sec:intro}
The coherent elastic neutrino-nucleus scattering (CE$\nu$NS) process
has been recently observed by the COHERENT experiment
\cite{Akimov:2017ade}, more than 40 years after its first theoretical
description \cite{Freedman:1973yd}.  Compared to other neutrino
processes at energies below 100 MeV, CE$\nu$NS has a large cross
section with a value of order $10^{-39}\,\text{cm}^2$, due to the
enhancement induced by the square of the number of neutrons in the
nucleus. However, despite these large values the CE$\nu$NS eluded
experimental detection for years due to the complicated measurement of
the weak nuclear recoil energies ($\sim$ few keV) produced in the
interaction. Its measurement became possible thanks to the development
of ultra-sensitive technology in other experimental searches namely,
rare decays and weakly interacting massive particle dark matter (DM) \cite{Akimov:2015nza}.

CE$\nu$NS occurs when the de Broglie wavelength of the scattering
process is larger than the nuclear radius ($\lambda=h/q\gtrsim R_N$,
where $q$ refers to the exchanged momentum), which for typical nuclei
translates into $q\lesssim 200\,$MeV. Accordingly, in $\nu-N$
scattering processes in which $q$ is sufficiently small the scattering
amplitudes on single nucleons add coherently and lead to an enhanced
cross section whose value depends upon the number of nucleons within
the nucleus.  In the standard model (SM) the CE$\nu$NS process is well understood and
it is determined by $Z$ boson exchange \cite{Freedman:1973yd}. It
receives contributions from vector and axial nuclear currents, with
the latter being---of course---relevant only for nuclei with spin
$J\neq 0$ \cite{Freedman:1977xn}. However, even in that case, it is
well known that the axial contribution is relevant only for light
nuclei \cite{Freedman:1977xn} and negligible for heavy ones, such as
Cs and I used in the COHERENT detector \cite{Akimov:2017ade}.

The COHERENT experiment uses neutrinos produced in the spallation
neutron source (SNS) at the Oak Ridge National Laboratory. The
spallation process starts with negatively charged Hydrogen ions H$^-$
which are accelerated at a LINAC. After being accelerated at
$\sim 0.9\,\text{c}$, the two electrons in the H$^-$ ions are stripped
off and the resulting protons are accumulated in a storage
ring. Spallation takes place when 60 Hz proton pulses hit a liquid
mercury fixed target. In that process not only neutrons but also pions
are produced from spallation. The neutrinos used by COHERENT are thus
generated by $\pi^+$ and $\mu^+$ decay resulting in prompt $\nu_\mu$
and delayed $\bar\nu_\mu$ and $\nu_e$ \cite{Akimov:2017ade}. The
detection of the CE$\nu$NS process is performed using a low-background
14.6 kg sodium-doped CsI (CsI[Na]) detector which relies on
scintillation for the detection of nuclear recoils induced by the
neutrino-nucleus scattering process.

Prospects for further and more precise measurements of the CE$\nu$NS
process include COHERENT upgrades. Starting with a CsI[Na]
scintillator detector, the collaboration plans upgrades involving
p-type point-contact Germanium and two-phase liquid Xe detectors
\cite{Akimov:2015nza}. The CENNS experiment is a proposal that aims at
using the Fermilab far-off-axis Booster Neutrino Beam, for which
$E_\nu\lesssim 50\,$MeV \cite{Brice:2013fwa}. In addition, experiments
using reactor anti-neutrinos with typical energies below $\sim 8\,$MeV
are also planned. They include: A proposal for an ultra-low-energy
Germanium detector located at the Kuo-Sheng Reactor Laboratory in
Taiwan, with a capability for measuring nuclear recoil energies down
to $0.1-0.2$ keV \cite{Wong:2008vk}; CONNIE, a solid state-based
detector using antineutrinos generated in the ``Almirante Alvaro
Alberto'' nuclear power plant in Rio de Janeiro, Brazil
\cite{Aguilar-Arevalo:2016qen}; CONUS a low-energy Germanium detector
which uses anti-neutrinos produced at a nuclear power plant in
Brokdorf, Germany \cite{conus}.

Measurements of the CE$\nu$NS process open a wide spectrum of physics
opportunities.  For example, they allow to constrain electroweak
parameters such as the weak mixing angle
\cite{Kosmas:2017tsq,Canas:2018rng}. They allow as well the study of
neutrino electromagnetic properties such as its charge radius or
magnetic dipole moment \cite{Vogel:1989iv}, which if present and
sufficiently sizable can affect---to a certain extent---the energy
recoil spectrum and the expected number of scattering events
\cite{Scholberg:2005qs,Kosmas:2017tsq}. They further provide a rich
avenue for testing the presence of more generic beyond the standard
model (BSM) physics, such as neutrino-quark non-standard interactions
(NSI) \cite{Barranco:2005yy,Scholberg:2005qs}, light mediators
associated with new $U(1)$ gauge symmetries or with extended scalar
sectors and involving order keV-MeV scalars
\cite{Shoemaker:2017lzs,Farzan:2018gtr,Denton:2018xmq,Billard:2018jnl}.
Moreover, facilities optimized for the study of the CE$\nu$NS process
offer a potential way to observe neutrinos from supernov\ae
\cite{Freedman:1977xn,Horowitz:2003cz}, measure the neutron part of
nuclear form factors \cite{Patton:2012jr} and test the presence of
sterile neutrinos \cite{Anderson:2012pn}. In addition, CE$\nu$NS
interactions share the same detectable signature (low-energy recoiling
nuclei) of galactic DM scattering off target nuclei. Hence, a precise
understanding of the CE$\nu$NS process is of paramount importance for
near-future DM direct detection searches, which will be subject to
irreducible neutrino backgrounds (solar neutrinos in the short-term)
\cite{Billard:2013qya,Essig:2018tss}.  New physics contributions can
potentially change the impact that such a background will have on the
capability of multi-ton scale DM detectors
\cite{Dutta:2017nht,AristizabalSierra:2017joc,Gonzalez-Garcia:2018dep,
  Papoulias:2018uzy}. So, understanding at which extent the new
physics can impact the CE$\nu$NS process becomes crucial.

In the wake of the observation of the CE$\nu$NS process by COHERENT,
various analyses have been considered. Soon after the data release,
ref. \cite{Coloma:2017ncl} studied the capability of the interplay
between neutrino oscillation and COHERENT data to rule out the
so-called Large Mixing Angle Dark (LMA-D) solution to neutrino mixing
parameters \cite{Coloma:2017ncl}, which arises in the presence of
NSI. COHERENT data allows, to a certain extent, to remove the
parameter degeneracy and to rule out the LMA-D solution at the
$\sim 3\sigma$ level. Ref. \cite{Liao:2017uzy} studied the constraints
implied by data on various BSM scenarios, including neutrino-quark NSI
and light vector mediators. These results show that COHERENT data
still allow for certain BSM scenarios to sizeably contribute to the
CE$\nu$NS cross section. Ref. \cite{Cadeddu:2017etk}, instead, focused
on nuclear aspects of the data and calculated for the first time the
root-mean-square neutron radius. More recently,
ref. \cite{Kosmas:2017tsq} carried out an analysis including NSI,
light vector and scalars (order MeV mediators) and electromagnetic
properties, with conclusions similar to those in
ref. \cite{Liao:2017uzy}. Other analyses have included DM properties,
e.g. constraints on dark-photon portal parameters, in which case
COHERENT places the most stringent bounds for
$m_\text{DM}\lesssim 30\,$MeV \cite{Ge:2017mcq}. This clearly shows
the capability of the CE$\nu$NS process as a tool for testing new
physics in the low-energy domain.

At the effective leading-order level, neutrinos couple to quarks
through dimension-six operators. In the SM both
vector and axial quark currents are present. Contributions from heavy
new physics can be parameterized by a larger set of couplings subject
only to the condition of Lorentz invariance. The most studied case of
such parameterization corresponds to neutrino NSI
\cite{Wolfenstein:1977ue}, where the couplings have a SM-like
structure, but are controlled by free parameters that ``measure'' the
relative strength of the new interaction to the Fermi interaction
($G_F$).  Neutrino NSI, however, are a subset of a whole set of
interactions which include scalar, pseudoscalar, vector, axial and
tensor couplings, which we refer to as \textit{neutrino generalized
  interactions} (NGI). They may emerge in BSM scenarios in which
e.g. neutrinos couple to heavy scalars \cite{Barger:2007im} or in
models where neutrinos have non-vanishing electromagnetic couplings
\cite{Lindner:2017uvt}. If we were to consider such scenarios
additional constraints from the charged lepton sector should be
accounted for\footnote{This is the case in gauge-invariant effective
  formulations of neutrino NSI. Additional constraints include limits
  on charged lepton flavor violating processes, universality of the
  $W^\pm$ gauge couplings and $e^+e^-\bar qq$ contact interactions
  \cite{Wise:2014oea}. Thus, in the case of NGI constraints of this
  type should apply as well.}, but here we do not consider this
possibility and rather stick from the very beginning to non-gauge
invariant dimension six operators. In doing so, we then place
constraints on the new effective couplings by requiring consistency
with the COHERENT measurement.  In our analysis we focus on the
leading contributions, which means that we do not consider
pseudoscalar nor axial quark currents. These are spin-dependent
interactions that---as in the SM case---lead to suppressed
contributions. It is worth emphasizing that our analysis is
complementary to those presented in
refs. \cite{Liao:2017uzy,Kosmas:2017tsq} and extends upon these
studies by including scalar effective interactions, crossed Lorentz
structures and simultaneous presence of different nuclear currents.

The rest of this paper is organized as follows. In sec. \ref{sec:gen}
we provide a short overview of the COHERENT experiment and discuss the
definitions and conventions used to perform our analysis, such as
theoretical neutrino fluxes and calculation of number of events.  We
also define the binned $\chi^2$ function and the different measured
quantities that are involved. In sec. \ref{sec:generalized} we present
the parametrization for NGI starting with neutrino-quark interactions
and ending up with neutrino-nucleus couplings. We provide relations
between the quark and nucleus couplings for scalar, vector and tensor
currents. In sec. \ref{sec:constraints} we present our results for the
differential cross section and the constraints implied on the
effective neutrino-quark parameters by COHERENT data. Finally, in
sec. \ref{sec:concl} we summarize and present our conclusions. In
appendix~\ref{sec:explicit-cal} we provide details of the cross
section calculation for neutrinos and anti-neutrinos in the
zero-momentum limit, including the full set of generalized
interactions for spin$-1/2$ nuclei.
\section{CE$\nu$NS signal rate at COHERENT}
\label{sec:gen}
COHERENT uses neutrinos produced in the Spallation Neutron Source at
the Oak Ridge National Laboratory \cite{Akimov:2017ade}.  The
interaction of a pulsed proton ($\sim 1\,\text{GeV}$) beam with a
fixed mercury target produces neutrons from spallation and a
substantial amount of low-energy neutrinos, which stem from the decay
of stopped pions and muons, $\pi^+\to \mu^++\nu_\mu$ and
$\mu^+\to e^++\nu_e+\bar\nu_\mu$. Muon neutrinos---being the
by-products of a two-body decay---are monochromatic, and their energy
is determined by the pion and muon masses:
$E_{\nu_\mu}=(m_\pi^2-m_\mu^2)/2m_\pi\simeq 30\,$MeV.  Accordingly,
their energy distribution is given by \cite{Coloma:2017egw}
\begin{equation}
  \label{eq:muon-neutrino-distribution-fun}
  \mathcal{F}_{\nu_\mu}(E_{\nu_\mu})=
  \frac{2m_\pi}{m_\pi^2-m_\mu^2}\,
  \delta\left(1-\frac{2E_{\nu_\mu}m_\pi}{m_\pi^2-m_\mu^2}\right) \, .
\end{equation}
Electron neutrinos and muon anti-neutrinos instead feature continuous
spectra. Their energy distribution---normalized to one---can be read
off from the $\mu^+$ (unpolarized) differential rate, namely
\begin{equation}
  \label{eq:muon-differential-rate}
  \frac{d}{dE_X}\Gamma(\mu^+\to e^++\nu_e+\bar\nu_\mu)=
  \Gamma(\mu^+\to e^++\nu_e+\bar\nu_\mu)
  \,\mathcal{F}_{\nu_X}(E_{\nu_X})
  \qquad
  (X=\nu_e, \bar\nu_\mu)\ ,
\end{equation}
where the energy distribution functions are given by \cite{Coloma:2017egw}
\begin{align}
  \label{eq:moun-dis-Fun}
  \mathcal{F}_{\nu_e}(E_{\nu_e})&=\frac{192}{m_\mu}
  \left(\frac{E_{\nu_e}}{m_\mu}\right)^2
  \left(\frac{1}{2}-\frac{E_{\nu_e}}{m_\mu}\right)\ ,
  \nonumber\\
  \mathcal{F}_{\bar\nu_\mu}(E_{\bar\nu_\mu})&=\frac{64}{m_\mu}
  \left(\frac{E_{\bar\nu_\mu}}{m_\mu}\right)^2
  \left(\frac{3}{4}-\frac{E_{\bar\nu_\mu}}{m_\mu}\right)\ ,
\end{align}
with the kinematic end point located at
$E_\nu=m_\mu/2\simeq 52.8\,$MeV.  The neutrino flux (per flavor) that
reaches the CsI[Na] detector,
$\phi_\alpha(E_{\nu_\alpha})$ ($\alpha=\nu_\mu, \bar\nu_\mu, \nu_e$),
is then determined by the energy distribution functions in 
eqs.~(\ref{eq:muon-neutrino-distribution-fun})-(\ref{eq:moun-dis-Fun})
times the total number of neutrinos per each flavor, $\mathcal{N}$. The
latter is fixed by the number of neutrinos produced per proton
collision ($r=0.08$ per flavor), the distance from the source to the
detector ($L=19.3\,$m) and the number of protons-on-target (POT,
$n_\text{POT}$). For the 308.1 live-days of neutrino production,
$n_\text{POT}=1.76\times 10^{23}$ \cite{Akimov:2017ade}. Thus, since
neutrinos are isotropically produced,
$\mathcal{N}=r\times \frac{n_\text{POT}}{4 \pi L^2}$ and
\begin{equation}
  \label{eq:fluxes}
  \phi_X(E_X)=\mathcal{N}\,\mathcal{F}_X(E_X)\qquad
  (X=\nu_\mu, \bar\nu_\mu, \nu_e)\ .
\end{equation}
The COHERENT detector consists of $m_\text{det}=14.6\,$kg of CsI[Na],
where the sodium dopant is present with a fractional mass of
$10^{-5}-10^{-4}$ and so it does not play any substantial r\^ole as a
target. Notice also that since $A_\text{Cs}\simeq A_\text{I}$, both Cs
and I yield approximately the same nuclear response. The number of
target nuclei is therefore given by
$n_N=\frac{2 m_\text{det}}{m_\text{CsI}} \times N_A$~\cite{Liao:2017uzy}, where
$m_\text{CsI}=2.598\times 10^{-1}\,$kg/mol is the CsI molar mass and
$N_A$ is the Avogadro number.

For a given flavor $\alpha$ and taking into account both the Cs and I
nuclei, the expected number of events in the $i$-th recoil energy bin
reads
\begin{equation}
  \label{eq:diff-event-rate-nucleus-dependent}
  R_\alpha^i=n_N\sum_{a=\text{Cs},\text{I}}f_a
  \int_{E_{r_i}-\Delta E_{r_i}}^{E_{r_i}+\Delta E_{r_i}}dE_r
  \mathcal{A}(E_r)\,F^2(q_a)
  \int_{E_\nu^\text{min}}^{E_\nu^\text{max}}dE_\nu
  \phi_\alpha(E_\nu)\,\frac{d\sigma_\alpha^a}{dE_\nu}\ .
\end{equation}
Here $E_\nu^\text{min}=\sqrt{2m_{N_a}E_r}$ ($E_r$ refers to the
nuclear recoil energy and $m_{N_a}$ to the nucleus mass),
$E_\nu^\text{max}=m_\mu/2$ and $f_a$ are nuclear fractions:
$f_\text{Cs}=51\%$ and $f_\text{I}=49\%$. The observed number of
photoelectrons (PE) is related to the recoil energy through
\cite{Akimov:2017ade}
\begin{equation}
  \label{eq:recoil-energy-PE}
  n_\text{PE}=1.17\left(\frac{E_r}{\text{keV}}\right)\ .
\end{equation}
In terms of $n_\text{PE}$ the COHERENT signal covers 25 bins starting
from $n_\text{PE}=1$ and extending up to $n_\text{PE}=49$, with bin
size equal to 2 photoelectrons. Since the acceptance function vanishes
for $n_\text{PE} \leq 5$ (see below), the first three bins contain no
information on the scattering process. Furthermore, from
$n_\text{PE}\geq 31$ the relation between the number of photoelectrons
and the nuclear recoil energy in (\ref{eq:recoil-energy-PE}) does not
hold anymore. Thus, in our analysis we consider only 14 data bins,
from $n_\text{PE}=7$ to $n_\text{PE}=31$\footnote{Recently, COHERENT
  has released data and detailed information that enables independent
  analysis \cite{Akimov:2018vzs}. In this release the acceptance
  function does cover the bin at $n_\text{PE}=5$. In our analysis
  however we use eq.~(\ref{eq:acceptance}), since the uncertainties in
  our calculation introduced by not considering this bin are small
  compared to e.g. uncertainties related with the choice of nuclear
  form factor ($\pm 5\%$) or neutrino fluxes ($\pm 10\%$)
  \cite{Akimov:2017ade}.}, assuming that at $n_\text{PE}=31$
eq. (\ref{eq:recoil-energy-PE}) is still valid (excluding this bin has
no significant impact in our results). In terms of $n_\text{PE}$ the
recoil energy integration limits are $(n_\text{PE}\mp 1)/1.17$.  In
our calculation we employ the nuclear Helm form factor (see discussion
in sec. \ref{sec:constraints}):
\begin{equation}
  \label{eq:helm}
  F(q_a)=3\frac{j_1(q_ar_n)}{q_ar_n}e^{q_a^2s^2/2}\ ,
\end{equation}
where $j_1(x)$ is the order-one spherical Bessel function,
$q_a=6.92\times 10^{-3}\sqrt{A_aE_r}\,\text{fm}^{-1}$ and the
effective nuclear radius is given by
$r_n=(c^2+7\pi^2a^2/3-5s^2)^{1/2}$, with $s=0.9\,$fm, $a=0.52\,$fm and
$c=(1.23A_a^{1/3}-0.6)\,$fm \cite{Lewin:1995rx}. The acceptance
function $\mathcal{A}(x)$ is given by \footnote{We thank Juan Collar
  and Bjorn Scholz from the COHERENT collaboration for giving us this
  information.}
\begin{equation}
  \label{eq:acceptance}
  \mathcal{A}(x)=\frac{k_1}{1+e^{-k_2(x-x_0)}}H(x-5)\ ,
\end{equation}
where $k_1=0.6655$, $k_2=0.4942$, $x_0=10.8507$ and $H$ is the
Heaviside function.

\begin{figure}
  \centering
  \includegraphics[scale=0.6]{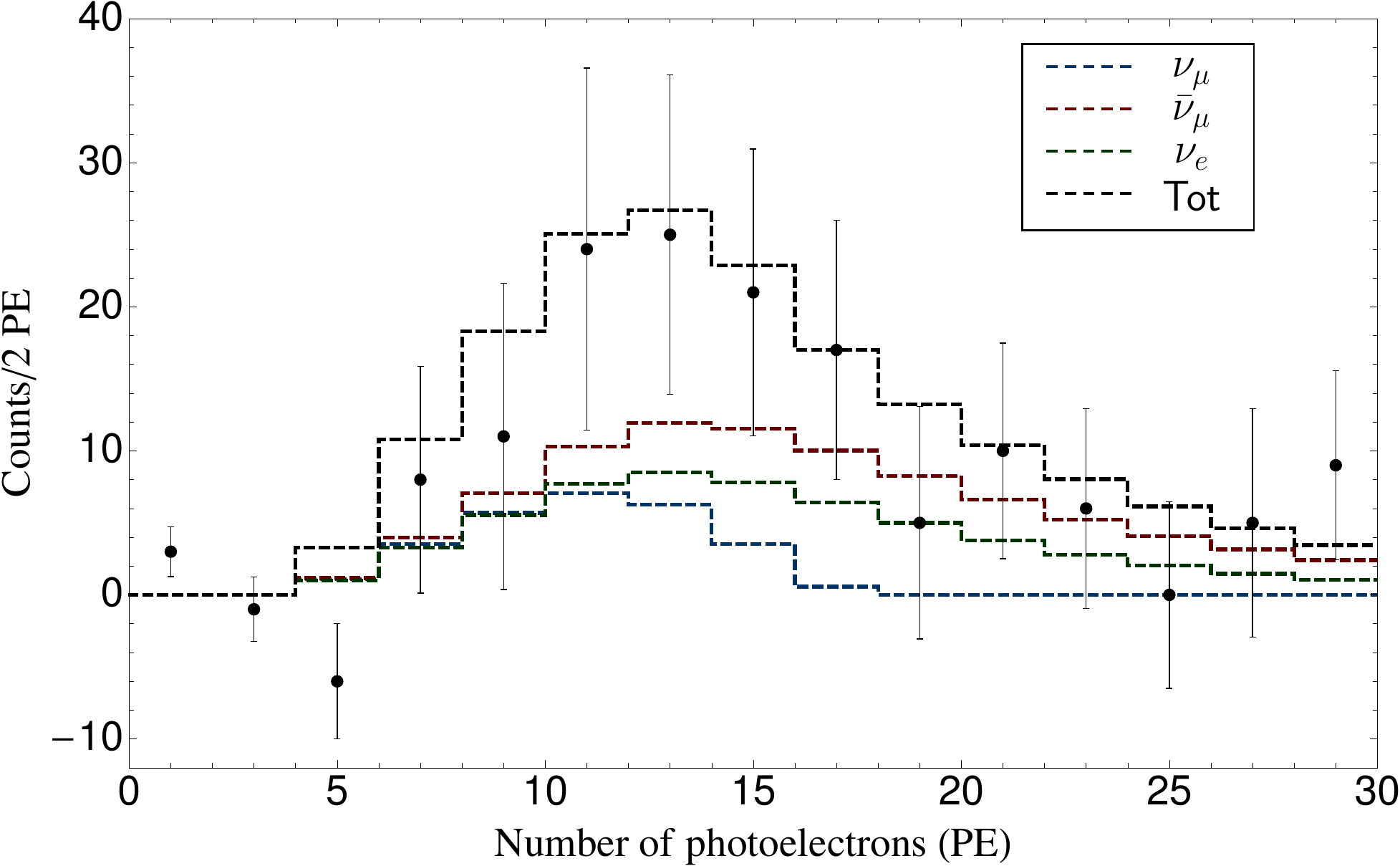}
  \caption{Number of expected CE$\nu$NS events as a function of
    photoelectrons in the SM. The points correspond to COHERENT data
    \cite{Akimov:2017ade} (with their error bars), while the colored
    histograms refer to the number of events from the three neutrino
    flavors produced in proton-Hg interactions. The largest (smallest)
    number of events is obtained from $\bar\nu_\mu$ ($\nu_\mu$) flux.}
  \label{fig:coherent-SM}
\end{figure}

The CE$\nu$NS differential cross section
$d\sigma_\alpha^a/dE_r$ depends on the nuclear target and in
BSM physics scenarios can be flavor dependent. In the SM
it arises from the neutral current vector and axial-vector
couplings \cite{Freedman:1973yd}, with the axial contribution terms
being subdominant \cite{Freedman:1977xn}. The leading contribution
can be written as follows:
\begin{equation}
  \label{eq:cneuNES-xsec}
  \frac{d\sigma^a}{dE_r}=\frac{G_F^2}{4\pi}m_{N_a}Q^2_{\text{SM},a}
  \left(1 - \frac{E_rm_{N_a}}{2E_\nu^2}\right)\,F^2(q^2)\ .
\end{equation}
Here $Q^2_\text{SM,a}=[Z_a(1-4\sin^2\theta_w)-N_a]^2\simeq N^2_a$,
thus showing that for heavy nuclei the CE$\nu$NS cross section is
largely enhanced. From
eqs. (\ref{eq:diff-event-rate-nucleus-dependent}), (\ref{eq:helm}),
(\ref{eq:acceptance}) and (\ref{eq:cneuNES-xsec}) we calculated the
number of CE$\nu$NS events predicted by the SM. The result is shown in
fig.~\ref{fig:coherent-SM}, for the different $\mathcal{F}_{X}$
separately (colored histograms) and for the total neutrino flux
(black histograms), together with the COHERENT data with their
corresponding uncertainties. As can be seen, these data closely
follows the SM prediction \cite{Akimov:2017ade}. However, due to the
still large uncertainties, sizable contributions from BSM physics can
be present and can therefore be constrained. As we have already
pointed out, since the release of the COHERENT result various BSM
scenarios have been analyzed.  They include neutrino effective NSI
\cite{Coloma:2017ncl,Liao:2017uzy}, NSI via light mediators
\cite{Liao:2017uzy,Kosmas:2017tsq,Farzan:2018gtr}, neutrino
four-fermion contact tensor interactions as well as electromagnetic
neutrino couplings \cite{Kosmas:2017tsq}.

\begin{figure}
  \centering
  \includegraphics[scale=0.3]{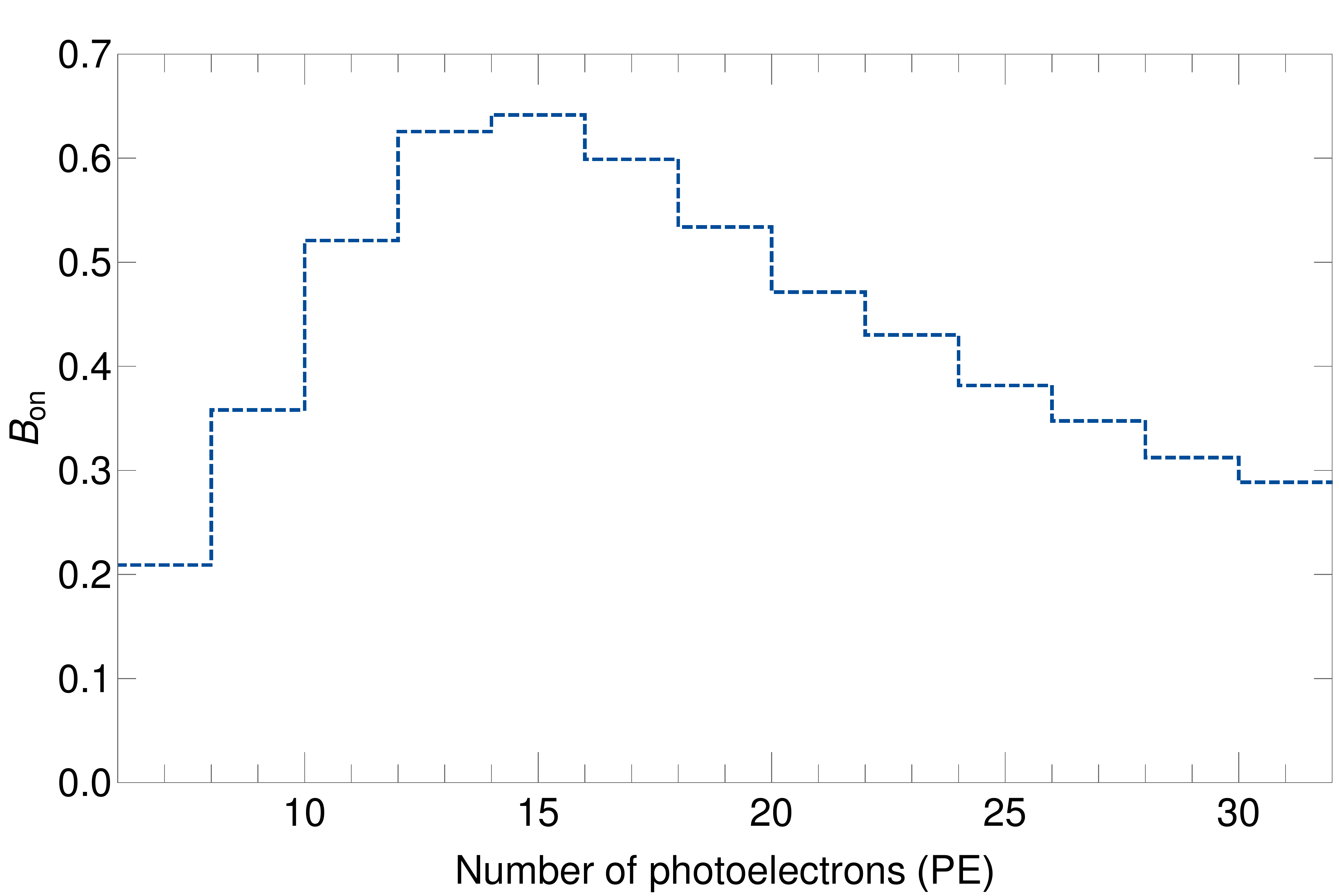}
  \caption{Beam-on background from prompt neutrons as a function of
    the number of photoelectrons $n_\text{PE}$. It follows from the prompt neutron
    probability distribution function and it is weighted by the energy
    delivered during the 308.1 live-days of neutrino production,
    $7.48\,$GWhr \cite{Akimov:2018vzs}. 
    Only PE bins considered in our analysis ($7 \leq n_\text{PE} \leq 31$)
    are shown.}
  \label{fig:Bon-per-bin}
\end{figure}

To constrain new physics contributions with COHERENT data, we use the
following least-squares function\footnote{Note that we consider a
  spectral definition for the $\chi^2$ function. Differently from the
  COHERENT collaboration, which performed an analysis of
  neutrino-quark NSI treating the measurement as a single-bin counting
  experiment~\cite{Akimov:2017ade}.}
\begin{equation}
  \label{eq:chiSq}
  \chi^2=\sum_{i=4}^{16}\left(\frac{N^\text{meas}_i
    -(1+\alpha)N^\text{NGI}_i(\mathcal{P})
    -(1+\beta)B^\text{on}_i}{\sigma_i}\right)^2
  + \left(\frac{\alpha}{\sigma_\alpha}\right)^2
  + \left(\frac{\beta}{\sigma_\beta}\right)^2\ ,
\end{equation}
where $N^\text{meas}_i$ is the number of events measured in the $i$-th
bin, $N^\text{NGI}_i$ is the number of events predicted by the NGI
scenario (determined by the set of parameters $\mathcal{P}$),
$\sigma_i^2$ is the statistical uncertainty on the experimental data
in the $i-$th bin. The nuisance parameters $\alpha$ and $\beta$
account for uncertainties on the signal rate and on the
background. Their standard deviations are $\sigma_\alpha=0.28$ and
$\sigma_\beta=0.25$. The calculation of the function in
eq.~(\ref{eq:chiSq}) requires as well the beam-on background (per bin)
$B_i^\text{on}$, which is dominated by far by prompt neutrons
resulting from the SNS and which are able to penetrate the 19.3 m of
moderating material. Fig. \ref{fig:Bon-per-bin} shows the distribution
for $B_i^\text{on}$, obtained from the prompt neutron probability
distribution function and weighted by the energy delivered during the
308.1 live-days of neutrino production, 7.48 GWhr
\cite{Akimov:2018vzs} \footnote{We thank Grayson Rich from the
  COHERENT collaboration for providing us this information prior to
  its release in \cite{Akimov:2018vzs} and for instructing us on its
  use.}.
%%%%%%%%%%%%%%%%%%%%%%%%%%%%%%%%%%%%%%%
\section{Neutrino generalized interactions}
\label{sec:generalized}
Most studies of the CE$\nu$NS  process in the presence of
new physics are done assuming neutrino NSI \cite{Wolfenstein:1977ue},
which are determined by the following four-fermion effective operator
\begin{equation}
  \label{eq:NSI}
  \mathcal{L}_{\text{eff}}^\text{NSI}=-\sqrt{2}G_F\sum_{q=u,d}\,
  \bar\nu_i\gamma^\mu(1-\gamma_5)\nu_j
  \,\bar q\left(\epsilon_{ij}^{qV}+\gamma_5\epsilon_{ij}^{qA}\right)q\ .
\end{equation}
Here $\epsilon_{ij}^{q(V,A)}$ are free parameters which are
constrained by neutrino oscillation and neutrino scattering data and
$q=u,d$ quarks
\cite{Coloma:2017egw,Coloma:2017ncl,Escrihuela:2009up,Bolanos:2008km}
(see also ref. \cite{Farzan:2017xzy} for a review). These couplings
parameterize the strength of the new interactions (relative to
$G_F$). The operator in eq.~(\ref{eq:NSI}) is actually more general
and encodes other interactions. For example, it describes as well an
effective theory involving operators such as
$(\bar\nu_{L_i}\,q_R)\,(\bar q_R\,\nu_{L_i})$, as can be checked by
Fierz rearrangement of the fermion fields.  Nevertheless,
eq.~(\ref{eq:NSI}) is not the most general effective $\nu-q$ operator.
A more general treatment is possible by considering all Lorentz
invariant non-derivative interactions of neutrinos with first
generation quarks, namely (we use the notation employed in
\cite{Lindner:2016wff})
\begin{equation}
  \label{eq:general-eff}
  \mathcal{L}_{\text{eff}}^{\rm NGI}=\frac{G_F}{\sqrt{2}}\sum_X\,
  \bar\nu\Gamma^X\nu
  \,\bar q\Gamma_X\left(C^q_X + i\gamma_5\,D_X^q\right)q\ .
\end{equation}
Here
$\Gamma^X=\{\mathbb{I},i\gamma^5,
\gamma^\mu,\gamma^\mu\gamma^5,\sigma^{\mu\nu}\}$,
where $\sigma^{\mu\nu}=i[\gamma^\mu,\gamma^\nu]/2$ and without loss of
generality the parameters $C^q_X$ and $D^q_X$ are real
\cite{Lindner:2016wff}. As in the NSI case, they ``measure'' the
relative strength of the new physics and so their size is of order
$(\sqrt{2}/G_F)(g_X^2/m_X^2)$, where $m_X$ is the mass of the exchanged particle 
and $ g_X$ the coupling constant. Due to the quark axial current term,
these interactions include diagonal and non-diagonal Lorentz
structures. For example, $\Gamma^P$ involves pseudoscalar-pseudoscalar
as well as pseudoscalar-scalar neutrino-quark couplings.

Among the NGI, those that give the most relevant effect, in the sense
that can sizeably diminish/exceed the SM contribution, do not involve
nuclear spin. Indeed, effective couplings for nuclear spin-dependent
interactions are determined by a sum over spin-up and spin-down
nucleons, $Z_\uparrow -Z_\downarrow$ and $N_\uparrow -N_\downarrow$
(for proton and neutrons respectively). Therefore they are suppressed
for all nuclei except for light ones \cite{Freedman:1977xn}. Since our
analysis involves heavy CsI nuclei we then drop the pseudoscalar and
axial quark currents and we only keep scalar, vector and tensor quark
currents. It is worth emphasizing that with this choice only the
parameters $\mathcal{P}=\{C_S^{q}$, $D_P^{q}$, $C_V^{q}$, $D_A^{q}$
and $C_T^{q}\}$ can be constrained.

To compute the CE$\nu$NS cross section induced by the NGI we assume a
fermion nuclear ground state with spin $J=1/2$. This is motivated by
the fact that nuclear matrix elements for nucleonic currents can in
this case be borrowed from nucleon matrix elements for quark
currents. Of course with such procedure one has to bear in mind that
the corresponding nuclear form factors are different. In our case all
the leading-order decompositions will involve the Helm form factor
given in eq.~(\ref{eq:helm})\footnote{We are assuming that the proton
  and neutron form factors are equal and well described by the Helm
  form factor, $F_Z(q^2)=F_N(q^2)\simeq F(q^2)$. A more precise
  approach in which $F_Z(q^2)$ is described by the Fourier transform
  of the symmetrized Fermi distribution and $F_N(q^2)$ by the Helm
  form factor could be adopted \cite{Cadeddu:2017etk}. However, given
  the uncertainties of COHERENT data our description is precise
  enough.}. This is somehow expected given that after dropping the
pseudoscalar and axial quark currents the remaining interactions
become spin-independent and so they add coherently on the
nucleons\footnote{This is what one finds in DM direct detection
  analyses: all nuclear interactions but the pseudoscalar and axial
  add coherently. Accordingly, apart from these two cases, the
  corresponding cross sections involve only the Helm form factor (see
  e.g. \cite{Lewin:1995rx,Fitzpatrick:2012ix}).}.

To determine the effective neutrino-nuclear Lagrangian, from which we
next calculate the cross section in the zero-momentum transfer limit,
we start with the quark currents and we end up with nuclear currents
following the procedure
\begin{equation}
  \label{eq:op-steps}
  \mathcal{O}_q\xrightarrow{\text{step (I)}} \mathcal{O}_{\mathcal{n}}
  \xrightarrow{\text{step (II)}}
  \mathcal{O}_N\ ,
\end{equation}
where $\mathcal{O}_{q,\mathcal{n},N}$ refer to quark, nucleon
($\mathcal{n}=p,n$) and nuclear operators, respectively. For step (I)
one calculates quark currents in nucleons according to (see
e.g. \cite{Dent:2015zpa,DelNobile:2013sia})
\begin{align}
  \label{eq:quark-to-nucleon-scalar}
  \langle\mathcal{n}(p_f)|\bar q\,q|\mathcal{n}(p_i)\rangle &=
  \frac{m_\mathcal{n}}{m_q}\,f_{T_q}\,
  \bar{\mathcal{n}}\,\mathcal{n}\ ,\nonumber\\
  %\label{eq:quark-to-nucleon-vector}
  \langle\mathcal{n}(p_f)|\bar q\gamma^\mu q|\mathcal{n}(p_i)\rangle&= 
  \mathcal{N}_q^{\mathcal{n}}\,
  \bar{\mathcal{n}}\,\gamma^\mu\,\mathcal{n}\ ,\,\nonumber\\[1mm]
  %\label{eq:quark-to-nucleon-tensor}
  \langle\mathcal{n}(p_f)|\bar q\,\sigma^{\mu\nu}\, q|\mathcal{n}(p_i)\rangle&=
  \delta_q^{\mathcal{n}}\,
  \bar{\mathcal{n}}\,\sigma^{\mu\nu}\,\mathcal{n}\ .
\end{align}
Here $p_i$ and $p_f$ refer to initial and final state nucleon
momenta. The scalar current receives contributions also from heavy
quarks ($q=c,b,t$), which are not of the form given in
eq.~\eqref{eq:quark-to-nucleon-scalar}. These contributions however
are suppressed by $m_{\mathcal{n}}/m_q$ and so we do not consider
them. Moreover, we neglect as well the contribution from strange
quarks and from gluons and we keep only first generation quarks.  For
vector currents, the coefficients $\mathcal{N}_q^{\mathcal{n}}$ can be
understood essentially as the number of quarks within the nucleon,
while for tensor currents $\delta_q^n$ represents a tensor charge. The
factors $f_{T_q}$ are related with the fraction of the nucleon mass
``carried'' by a particular quark flavor.  They are derived in chiral
perturbation theory from measurements of the $\pi-\mathcal{n}$ sigma
term \cite{Cheng:1988im}. The factors $\delta_q^n$ that we use here
are derived from an analysis based on data from azimuthal asymmetries
in semi-inclusive deep-inelastic scattering (DIS) and
$e^+e^-\to h_1 h_2 X$ processes \cite{Anselmino:2008jk}. More recent
values are given in
refs. \cite{Courtoy:2015haa,Goldstein:2014aja,Radici:2015mwa}. In our
calculation we use the numerical values
\cite{Jungman:1995df,Anselmino:2008jk}
\begin{alignat}{3}
  \label{eq:nucleon-factors}
  f_{T_u}^p=0.019\ ,\quad& f_{T_d}^p=0.041\ ,\quad&
  \delta^p_u=0.54\ ,\quad& \delta^p_d=-0.23\ ,
  \\
  f_{T_u}^n=0.023\ ,\quad& f_{T_d}^n=0.034\ ,\quad&
  \delta^n_u=-0.23\ ,\quad& \delta^n_d=0.54\ .
\end{alignat}

For step (II) one evaluates the correlators of nucleonic currents in
nuclei, which involve nuclear form factors and which can be written
following Lorentz invariance, namely
\begin{align}
  \label{eq:nucleon-to-nucleus-scalar}
  \langle N(k_2)|\bar{\mathcal{n}}\,\mathcal{n}|N(p_2)\rangle &=
  \bar N\,N F(q^2)\ ,\nonumber\\
  %\label{eq:nucleon-to-nucleus-vector}
  \langle N(k_2)|\bar{\mathcal{n}}\gamma^\mu \mathcal{n}|N(p_2)\rangle&=
  \bar N \left(
  \gamma^\mu F(q^2)
  +
  \frac{\sigma^{\mu\nu}q_\nu}{2m_N}F_1(q^2)
  \right)N\ ,\nonumber\\[1mm]
  %\label{eq:nucleon-to-nucleus-tensor}
  \langle N(k_2)|\bar{\mathcal{n}}\,\sigma^{\mu\nu}\, \mathcal{n}|N(p_2)\rangle&=
  \bar N\,
  \left(
  i\sigma^{\mu\nu}F(q^2) 
  - \frac{\gamma^\mu q^\nu - \gamma^\nu q^\mu}{2m_N}F_2(q^2)
  - \frac{K^\mu q^\nu - K^\nu q^\mu}{2m_N^2}F_3(q^2)
  \right)\,N\ ,
\end{align}
where the momenta of the incoming and outgoing nucleus, $p_2$ and
$k_2$, define the exchanged momentum $q=k_2-p_2$.  Some words are in
order regarding these decompositions. $F(q^2)$ refers to the Helm form
factor in eq.~\eqref{eq:helm} and it is in practice the only one
relevant at leading order. The magnetic moment term in the vector
current decomposition, as well as the second and the third terms in
the tensor decomposition, are suppressed by $\mathcal{O}(q/m_N)$
factors. Thus, keeping just the leading terms, step (II) can be
carried out and the neutrino-nucleus ($\nu-N$) effective Lagrangian
can be written
\begin{align}
  \label{eq:nu-N-Lag}
  \mathcal{L}_{\nu-N}\sim
   \sum_{X=S,V,T}\bar\nu\,\Gamma_X\nu\,\bar N\,C_X\,\Gamma_X\,N
   +
   \sum_{\substack{(X,Y)=(P,S),\\\qquad\,\,\,(A,V)}}
   \bar\nu\,\Gamma_X\nu\,\bar N\,iD_X\,\Gamma_Y\,N\ ,
\end{align}
where the coefficients $C_X$ and $D_X$ correspond to $\nu-N$ effective
couplings determined by the parameters $C_X^{q}$ and $D_X^{q}$ in
eq.~\eqref{eq:general-eff}. Notice that from eq.~\eqref{eq:nu-N-Lag}
we can calculate the zero-momentum cross section, while the full cross
section will involve the nuclear form factor which in turn will encode
the momentum dependence ($q^2$ dependence). The $\nu-N$ coefficients
are written as follows:
\begin{align}
  \label{eq:nuclear-coefficients}
  C_S&=Z\sum_{q=u,d}C_S^{(q)}\frac{m_p}{m_q}f^p_{T_q} 
  + (A-Z)\sum_{q=u,d}C_S^{(q)}\frac{m_n}{m_q}f^n_{T_q}\ ,
  \nonumber\\
  C_V&=Z\left(2 C^u_V + C^d_V\right) 
  + (A-Z)\left(C^u_V + 2C^d_V\right)\ ,
  \nonumber\\[2mm]
  C_T&=Z\left(\delta_u^p C^u_T + \delta_d^p C^d_T\right) 
  + (A-Z)\left(\delta_u^n C^u_T + \delta_d^n C^d_T\right)\ . 
\end{align}
The expression for $D_P$ is obtained from that of $C_S$ by trading
$C_S^{q}$ for $D_P^q$, while for $D_A$ from $C_V$ by trading $C_V^{q}$
for $D_A^q$. The relations in eq.~\eqref{eq:nuclear-coefficients}
allow to translate the constraints on the $\nu-N$ coefficients to the
parameters of the ``fundamental'' Lagrangian.

% ------------------
% Begin sub-section
% ------------------
\subsection{Neutrino oscillations versus neutrino scattering}
\label{sec:constraints-from-oscillations}
Before proceeding with the chi-square analysis, it is worth commenting
on which other processes may set constraints on the NGI and on the
range of validity of our results. As in the NSI case, interactions in
eq.~(\ref{eq:general-eff}) contribute---in principle---to forward
coherent scattering (order $G_F$ at $q^2=0$) and scattering processes
(order $G_F^2$ with $q\neq 0$). The former are responsible for matter
potentials in matter and are related to neutrino oscillation data,
while the latter include not only COHERENT but also DIS data from
CHARM and NuTeV \cite{Dorenbosch:1986tb,Zeller:2001hh}.

Matter potential induced by SM vector interactions in the Sun and in
the Earth are responsible for resonant neutrino flavor conversion
\cite{Wolfenstein:1977ue,Mikheev:1986gs,Mikheev:1986wj}. Accordingly,
new contributions to the vector current are subject to both
constraints, oscillation+scattering. This is indeed the case for
neutrino NSI, where it is found that the combined analysis of
oscillation+scattering data imply more stringent bounds
\cite{Coloma:2017egw,Esteban:2018ppq}. Scalar interactions couple
background fermions (nucleons) with different chiralities but same
helicity, and so they lead to helicity suppressed matter potentials
($m_\nu/\langle E_\nu\rangle$) \cite{Bergmann:1999rz}. Constraints
from neutrino oscillation data on these couplings are thus loose, if
existing at all. Transverse tensor interactions, instead, can induce a
sizable matter potential as they couple background fermions with
different chiralities and opposite helicities (longitudinal tensor
interactions are helicity suppressed as well). However, this tensor
matter potential is only relevant in a polarized medium and so it does
not sizeably affect neutrino propagation in the Sun or even in
supernov\ae \cite{Bergmann:1999rz}.

In the NSI case, DIS data places more severe bounds than COHERENT data
does \cite{Coloma:2017egw}. This should apply as well for the
remaining interactions in (\ref{eq:general-eff}). These limits however
do not apply for mediators whose masses are below the typical momentum
exchange in DIS processes, $\mathcal{O}(10\,\text{GeV})$. For
$m_X^2\ll q^2_\text{DIS}$, the relative value of the new contribution
is $\sigma_\text{BSM}/\sigma_\text{SM}\sim g_X^4/q^4/G_F^2$ and
amounts to $1\%$ for $g_X=10^{-2}$. The same parameter choice with
$m_X=10^2\,$MeV and evaluated at
$q_\text{COH}^2\simeq (10\,\text{MeV})^2$ gives
$\sigma_\text{BSM}/\sigma_\text{SM}\gg 1$. This means that for
mediator masses below $10^3\,$MeV DIS constraints can be evaded and
COHERENT bounds become dominant.

It follows that the constraints we derive here (see
sec.~\ref{sec:constraints}) are the most stringent (for all
interactions except the vector one), in scenarios where the mediator
mass is below $10^3\,$MeV. For heavier mediators, more severe limits
from DIS data may apply, but to the best of our knowledge such bounds
do not exist.
% -----------------------
% Begin section
% -----------------------
\section{Constraints from COHERENT data}
\label{sec:constraints}
To address the implications of COHERENT data on NGI, one has to
calculate the number of expected events for a certain parameter choice
according to eq.~\eqref{eq:diff-event-rate-nucleus-dependent}. This
requires the determination of the corresponding cross sections for
$\nu-N$ and $\bar \nu-N$ coherent scattering (the former has been
derived in \cite{Lindner:2016wff}).  Starting from the Lagrangian in
eq.~\eqref{eq:nu-N-Lag} we calculate the zero-momentum differential
cross section at leading order, i.e. neglecting
$\mathcal{O}\left(E^2_r/E^2_\nu\right)$ terms:
\begin{align}
  \label{eq:diff-x-sec}
  \frac{d\sigma^a(q^2=0)}{dE_r}=
  \frac{G_F^2}{4\pi}m_{N_a}N_a^2
  \left[\xi_S^2\,\frac{E_r}{E_r^\text{max}} 
  + \xi_V^2\left(1 - \frac{E_r}{E_r^\text{max}} - 
  \frac{E_r}{E_\nu}\right)
  +
  \xi_T^2
  \left(1 - \frac{E_r}{2E_r^\text{max}}-\frac{E_r}{E_\nu}\right)
  - R\frac{E_r}{E_\nu}
  \right]\ ,
\end{align}
the index $a$ denoting the target material. Details of the full
calculation, including pseudoscalar and axial quark currents, are
given in app. \ref{sec:explicit-cal}.  In the previous expression,
$E_r^\text{max}\simeq 2E_\nu^2/m_{N_a}$ and the following definitions
apply\footnote{These definitions slightly differ from what was found
  in ref. \cite{Lindner:2016wff}, where $\xi_V=(C_V-D_A)/N$.}
\begin{align}
  \label{eq:xi-definitions}
  \xi_S^2=\frac{C_S^2+D_P^2}{N^2}\ ,
  \quad
  \xi_V^2=\frac{C_V^2+D_A^2}{N^2}\ ,
  \quad
  \xi_T^2=8\frac{C_T^2}{N^2}\ ,
  \quad
  R=2\frac{C_SC_T}{N^2}\ .
\end{align}

The $\xi_X$ parameters defined in eq.~\eqref{eq:xi-definitions} depend
upon the nucleus, although for the sake of simplicity we have not
written this dependence explicitly.  The momentum-dependent cross
section is then obtained from eq.~(\ref{eq:diff-x-sec}) introducing
the nuclear form factor
\begin{equation}
  \label{eq:momentum-dependent-x-sec}
  \frac{d\sigma^a(q^2)}{dE_r}=\frac{d\sigma^a(q^2=0)}{dE_r}F(q^2)\ .
\end{equation}
Should an axial nuclear current be present, eq.~\eqref{eq:diff-x-sec}
would contain two additional terms, corresponding to the axial
contribution itself and to an interference term between the vector and
axial currents. This axial-vector interference term as well as the
last term in eq.~\eqref{eq:diff-x-sec} (proportional to $R$) are the
only two that come with opposite signs in the $\bar\nu-N$ and $\nu-N$
cross sections. In the former (latter) case we find that the
vector-axial interference term leads to constructive (destructive)
interference. If we neglect pseudoscalar and axial nuclear currents
then the neutrino and anti-neutrino elastic scattering cross sections
differ only in the term proportional to $R$, which turns out to be
relevant only if scalar and tensor interactions are simultaneously
present. This term leads to rather suppressed differences and
eventually the neutrino and anti-neutrino cross sections can be
considered equal.

In full generality, the parametrization introduced in
eq.~\eqref{eq:xi-definitions} must include the SM as well.  The SM
limit is recovered when all the couplings but $\xi_V=C_V$ are set to
zero and $\xi_V=C_V=1-(1-4\sin^2\theta_w)\times N/Z$. This
contribution is of course always present throughout our analysis, and
so from now on we will denote by $\xi_V$ the BSM contribution to the
vector current.  Note that the term proportional to $\xi_V$ has en
extra term, $E_r/E_\nu$, compared to the SM cross section for $J=0$,
eq. (\ref{eq:cneuNES-xsec}). This term is a consequence of the nuclear
ground state spin, $J=1/2$ \cite{Lindner:2016wff}. From
eqs. (\ref{eq:nuclear-coefficients}) and (\ref{eq:xi-definitions}) one
can see that even neglecting pseudoscalar and axial quark currents and
without considering lepton flavor dependent couplings, the full
problem involves 10 free parameters.  In order to technically simplify
the analysis, rather than considering the whole set, we stick to two
kinds of simplified benchmark scenarios which we will discuss in the
next subsections.
% ------------------
% Subsection one-P
% ------------------
\subsection{Single-parameter scenarios}
\label{sec:one-parameter}
\begin{table}
  \centering
  \renewcommand{\arraystretch}{1.4}
  \setlength{\tabcolsep}{0.9em}
  \begin{tabular}{|c|c|c|c|}\hline
    \textbf{Param} & \textbf{BFP value} & 
    $90\%$ CL & $99\%$ CL\\\hline\hline
    $\xi_S$ & $ 0 $ & $[-0.62,0.62] $ & 
    $[-1.065,1.065]$ \\\hline
    \multirow{2}{*}{$\xi_V$}& $-0.113 $ & $[-0.324, 0.224] $ & 
    $[-0.436, 0.67]$\\
    & $-1.764$ & $[-2.102,-1.554]$ & $[-2.545,-1.442]$\\\hline
    $\xi_T$ & $0 $ & $[-0.591,0.591]$ & 
    $ [-1.071,1.072]$ \\\hline
  \end{tabular}
  \caption{Best-fit-point value (second column), 90\% CL 
    ($\Delta \chi^2 < 2.71$, third column) 
    and 99\% CL ($\Delta \chi^2 < 6.63$, fourth column)
    ranges for the $\xi_X$ ($X=S, V, T$) parameters as defined
    in eq. (\ref{eq:xi-definitions}). From these results 
    one can then map to the fundamental neutrino-quark
    parameters using eq.~(\ref{eq:nuclear-coefficients}).
    See text for further details.}
  \label{tab:xi-x-parameters-single-parameter-case}
\end{table}
We start our analysis by considering the single-parameter case
parameterized in terms of the different $\xi_X$. These ``couplings''
are related to the neutrino-quark couplings of the ``fundamental''
Lagrangian through the relations derived in
eq.~(\ref{eq:nuclear-coefficients}). Thus, in reality, by
``single-parameter" scenarios we refer to the cases in which only one
interaction \textit{at the nuclear level} is present at a time. This,
however, does not mean that the analysis reduces to a single parameter
problem. Take for example the vector case. Vector nuclear currents
arise from either $\Gamma^V\,\Gamma^V$ or $\Gamma^A\,\Gamma^A$ Lorentz
structures, as can be seen by the definition of $\xi_V$.  As already
introduced in sec.~\ref{sec:generalized}, in our analysis we only
consider first generation quarks. Thus, in general, when considering
the case in which all $\xi_X$ vanish except for $\xi_V$, one is
eventually dealing with a four-parameter problem
($C_V^u, C_V^d, D_A^u$ and $D_A^d$).  Similarly, also the scalar
interaction involves four fundamental parameters
($C_S^u, C_S^d, D_P^u$ and $D_P^d$), encoded in $\xi_S$. The tensor
current instead depends only on two parameters at the quark level,
$C_T^u$ and $C_T^d$.

\begin{figure}[t]
  \centering
  \includegraphics[scale=0.45]{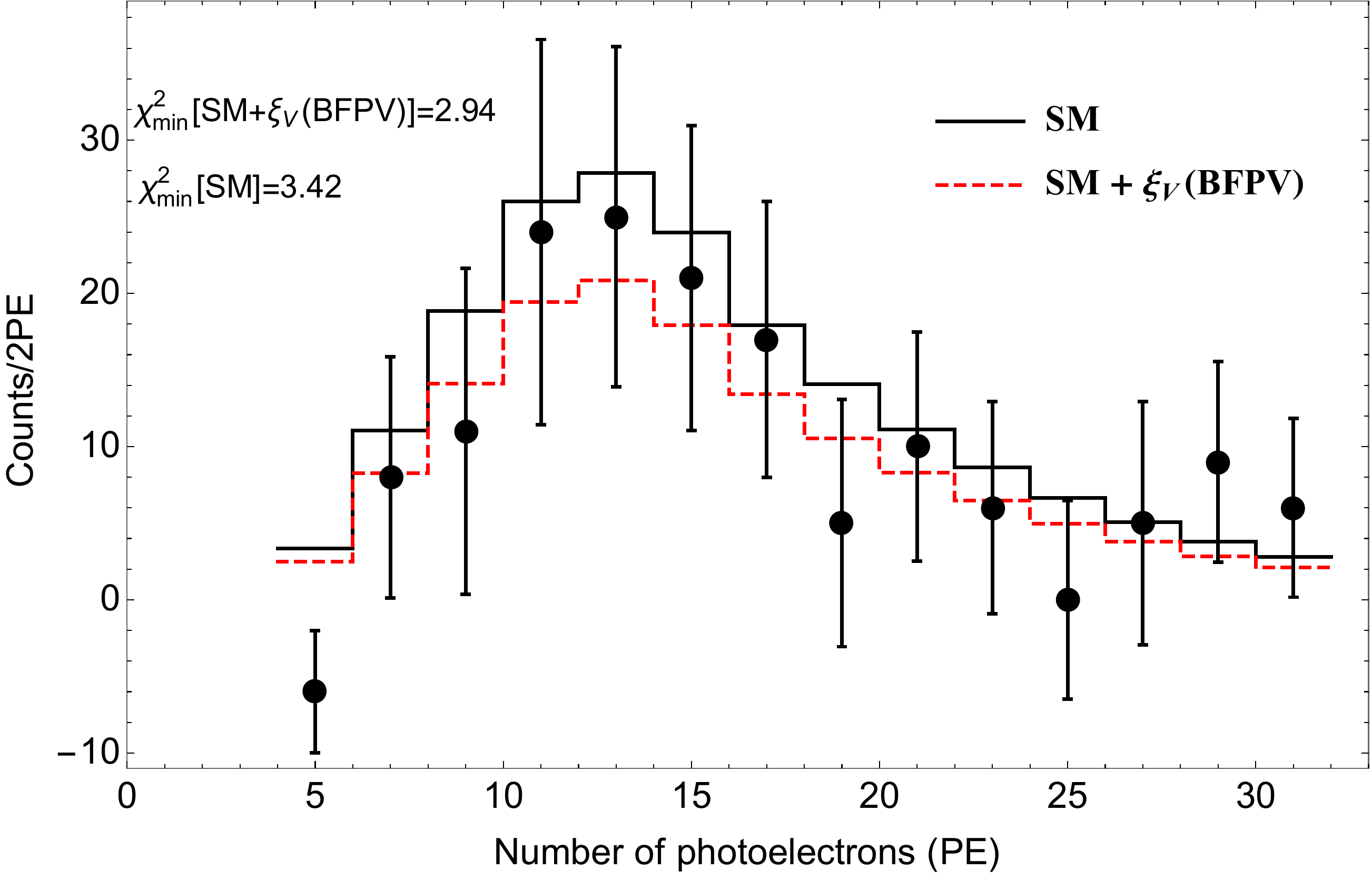}
  \caption{Number of expected CE$\nu$NS events as a function of the
    number of photoelectrons in the SM and in a BSM scenario with
    vector interactions (SM + $\xi_V$(BFPV)).  The points correspond to
    COHERENT data \cite{Akimov:2017ade}, while the black (red) colored
    histograms refer to the numbers of events from the three neutrino
    flavors in the SM (SM + $\xi_V$(BFPV)). See text for details.}
  \label{fig:csiV_num_events}
\end{figure}

To facilitate the numerical analysis, for the scalar and vector cases
we will consider only two neutrino-quark parameters at a time, for
which there are six possible choices: (1-i) $C_X^{q}\neq 0$ or
$D_X^{q}\neq 0$, (1-ii) $C_X^u$ and $D_X^d$ different from zero (or
$C_X^d, D_X^u\neq 0$), (1-iii) $C^u_X$ and $D^u_X$ different from zero
(or $C^d_X, D^d_X\neq 0$). Of these cases, (1-i) and (1-ii) lead to
the same constraints over the different parameters. Constraints
derived on $C_X^{q}$ apply directly on $D_X^{q}$ and those derived on
$C_X^u$ and $D_X^d$ on $C_X^d$ and $D_X^u$. Cases (1-iii) instead
result in different constraints over the different
couplings. Nevertheless, they should differ only by small values,
given that the differences between the up and down couplings and
masses are small (this is actually what is found in NSI analyses
\cite{Coloma:2017egw,Esteban:2018ppq}). We thus consider only the
first options in (1-i)-(1-iii) and $C^u_T$-$C^d_T$ for the tensor
interaction.

For all scenarios we fit the COHERENT data by minimizing the
least-squares function (eq.~\eqref{eq:chiSq}) over the systematic
nuisance parameters $\alpha$ and $\beta$, and then we calculate
$\Delta\chi^2 = \chi^2 -\chi^2_{\rm min}$. From this procedure we
obtain the $90\%$ and $99\%$ CL allowed ranges for each $\xi_X$. Our
results are shown in
tab.~\ref{tab:xi-x-parameters-single-parameter-case}.  Note that while
the best fit point values (BFPVs) for $\xi_S$ and $\xi_T$ are zero, an
additional vector current with $\xi_V = -0.113~(-1.764)$
(corresponding to the two minima of the $\Delta\chi^2(\xi_V)$
function) improves the COHERENT data fit. This is shown in
fig.~\ref{fig:csiV_num_events}, where the black (red) colored
histograms refer to the CE$\nu$NS number of events from the three
neutrino flavors in the SM (SM plus a vector NGI, with
$\xi_V = -0.113$).  Values for $\chi^2_{\rm min}$ in both cases are
also shown.

\begin{figure}[t]
  \centering
  \includegraphics[scale=0.3]{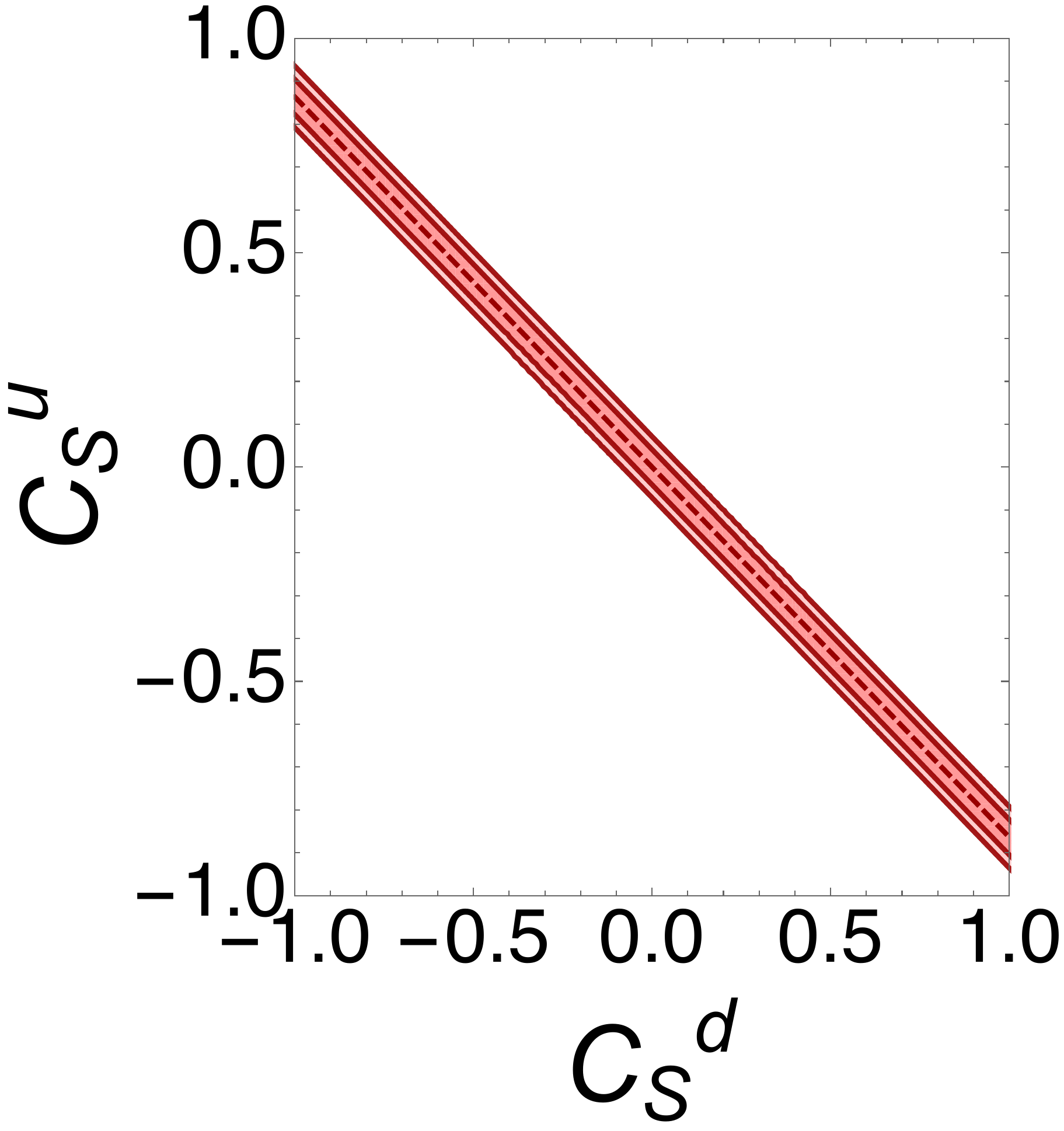}
  \hspace{1cm}
  \includegraphics[scale=0.285]{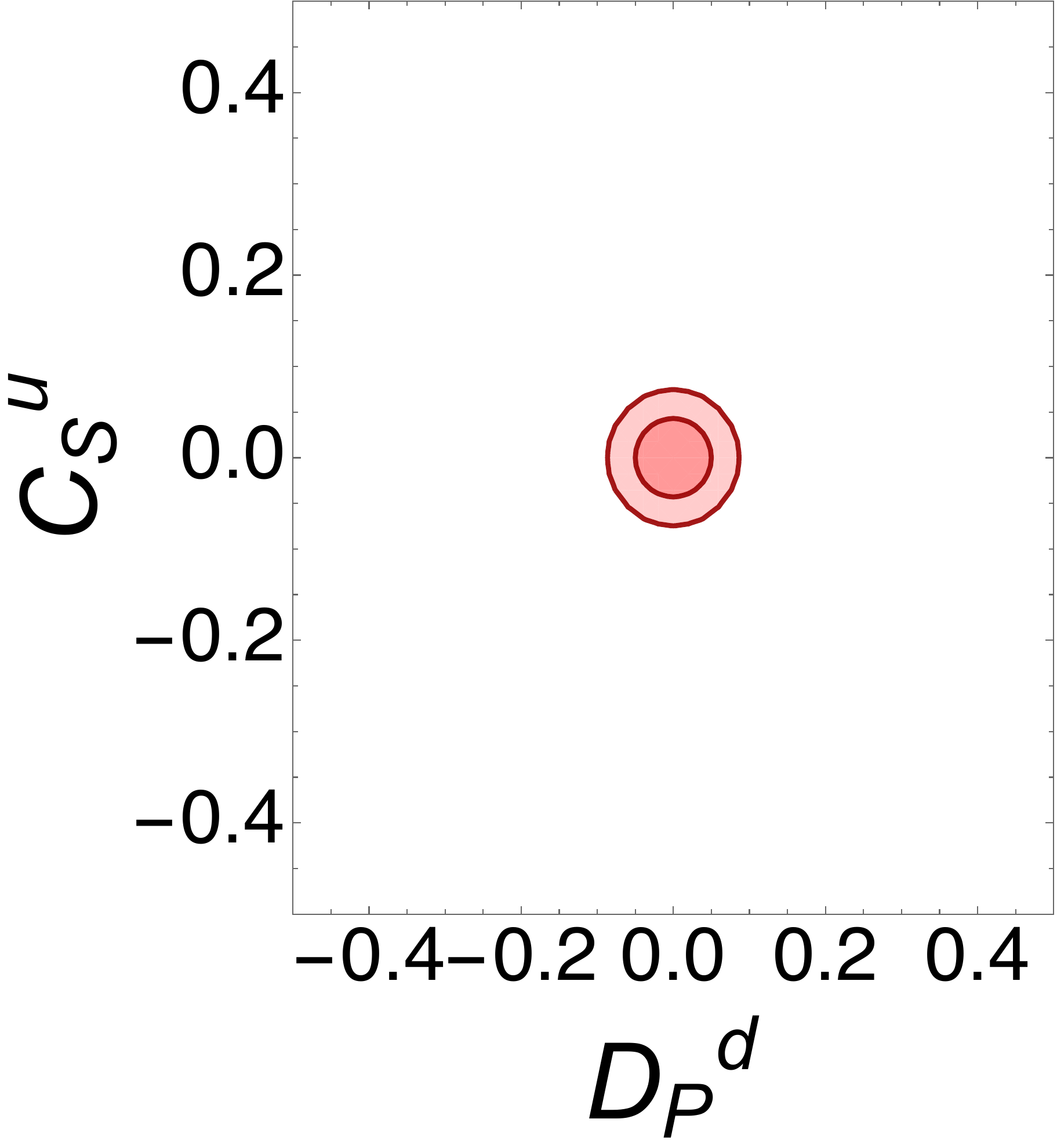}\\
  \includegraphics[scale=0.3]{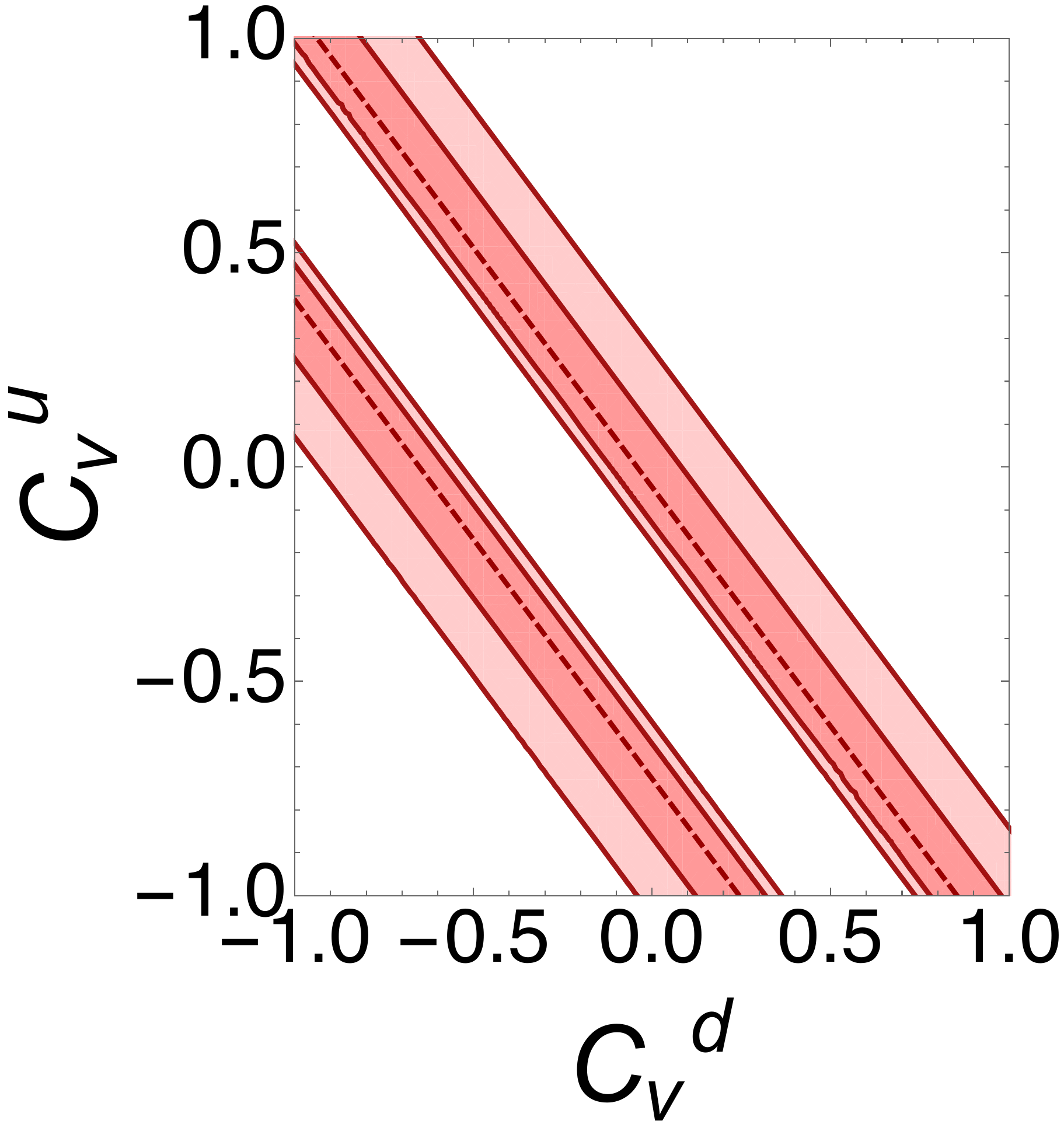}
  \hspace{1cm}
  \includegraphics[scale=0.285]{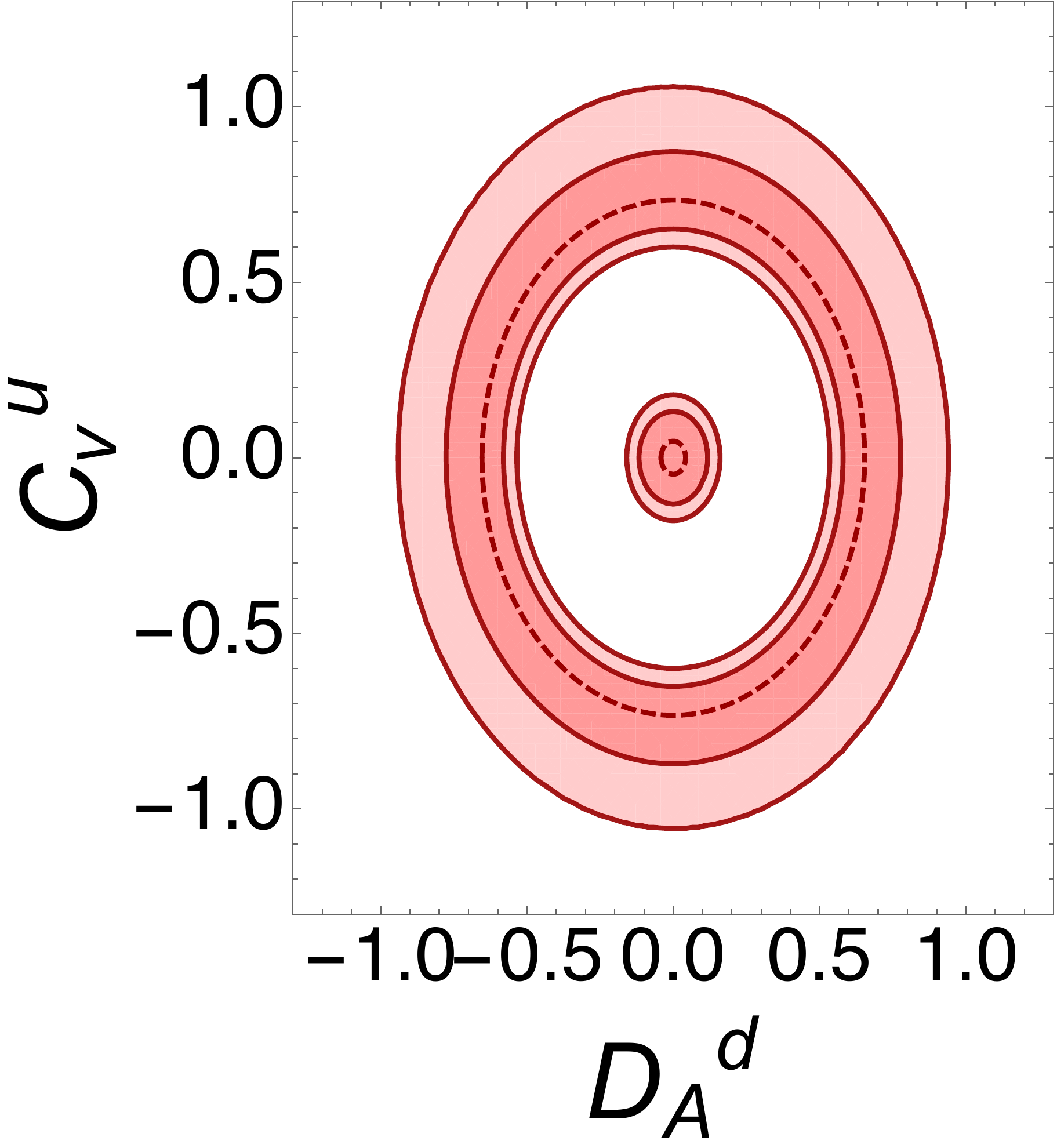}
  \caption{90\%CL ($\Delta \chi^2 < 2.71$, dark reddish) and 99\%CL
    ($\Delta \chi^2 < 6.63$, light reddish) allowed regions in the
    neutrino-quark couplings parameter space for scalar and vector
    interactions. Panels in the left column correspond to constraints
    for scenario (1-i), while those in the second column to scenario
    (1-ii). Results for scenario (1-iii) resemble those of scenario
    (1-ii) and so we do not display them. The dashed lines refer to
    the values determined by the $\xi_{S,V}$ BFPVs (see
    tab. \ref{tab:xi-x-parameters-single-parameter-case}). For $\xi_V$
    the $\chi^2$ function exhibits two minima and so for this case the
    result includes two non-overlapping regions.}
  \label{fig:scalar-vector-single-param}
\end{figure}

\begin{figure}[t]
  \centering
  \includegraphics[scale=0.3]{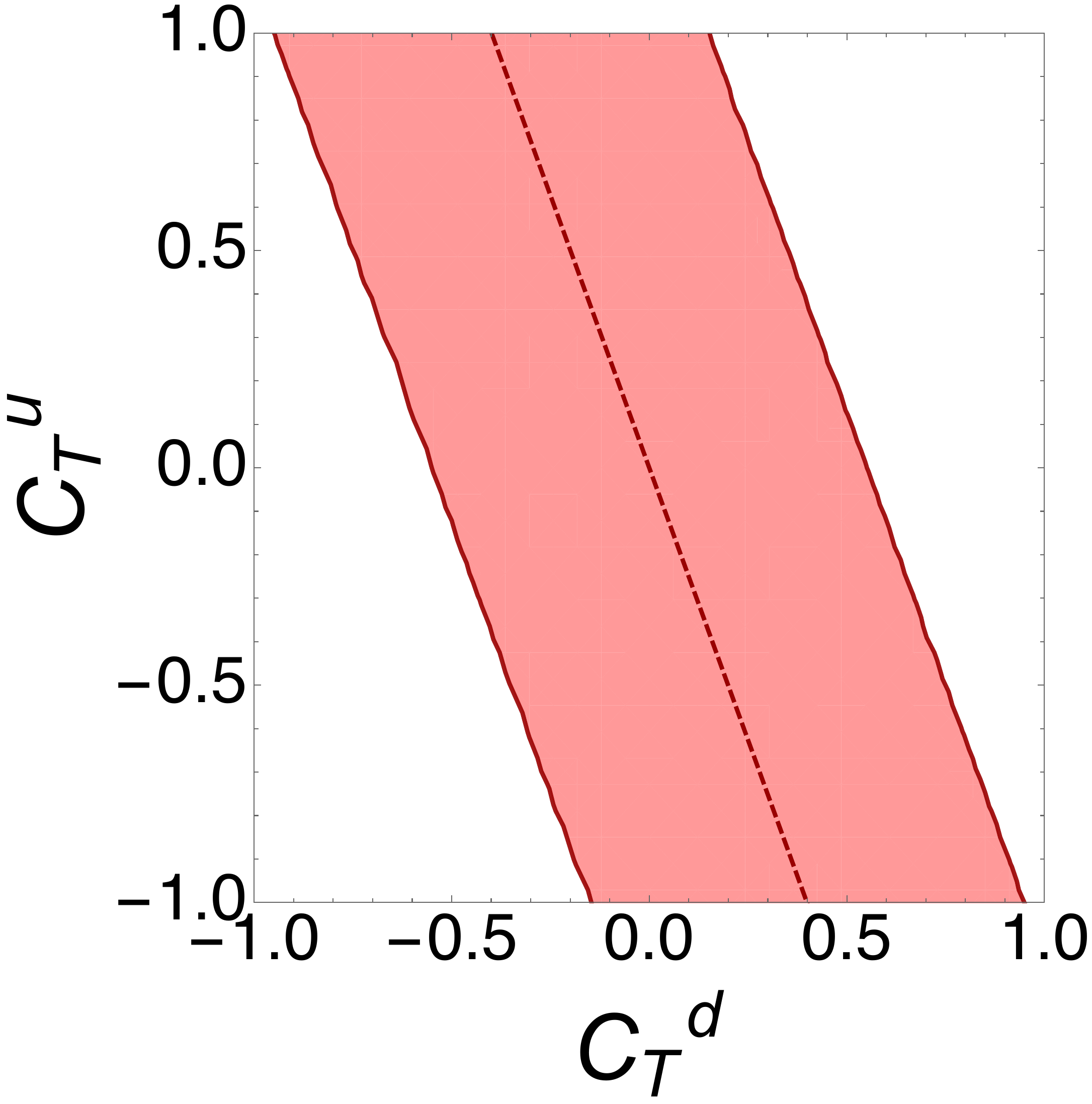}
  \caption{90\%CL allowed regions in the neutrino-quark couplings
    parameter space for tensor interactions. The dashed line refer to
    the value determined by the $\xi_T$ BFPV (see
    tab. \ref{tab:xi-x-parameters-single-parameter-case}).}
  \label{fig:tensor-single-param}
\end{figure}

The results in tab.~\ref{tab:xi-x-parameters-single-parameter-case}
can be translated into the ``fundamental'' neutrino-quark parameters
by using eq. (\ref{eq:nuclear-coefficients}). To do so one has to bear
in mind that although the number of events receives contributions from
Cs and I, the following simplification applies
$\xi_X^2\simeq \xi_{X_\text{Cs}}^2 + \xi_{X_\text{I}}^2\simeq
2\xi_{X_\text{Cs}}^2$.
As expected, given that $f_\text{Cs}\simeq f_\text{I}$,
$F^2(q_\text{Cs})\simeq F^2(q_\text{I})$ and
$m_{N_\text{Cs}}\simeq m_{N_\text{I}}$ (numerically we find
$\xi_{X_\text{I}}/\xi_{X_\text{Cs}}\simeq 0.95$ for all $X$).  We then
derive the allowed 90\%, 99\% CL regions for the quark parameters for
scenarios (1-i)-(1-iii) and for the tensor case in terms of
$C_T^u-C_T^d$. Fig. \ref{fig:scalar-vector-single-param} shows the
result for the scalar and vector interactions for scenarios (1-i) and
(1-ii) (results for scenario (1-iii) closely resemble those from
(1-ii) and so we do not display them), while
fig. \ref{fig:tensor-single-param} for the tensor couplings. It is
worth emphasizing that the $C^q_X, D^q_X$ couplings appearing in the
different panels in fig. \ref{fig:scalar-vector-single-param} are not
independent, as can be seen from the translation of the $\xi_X$
parameters into the $C^q_X$ couplings,
eq.~\eqref{eq:nuclear-coefficients}. Hence, even if we display the CL
contour regions in two quark parameter planes, since the initial
$\chi^2$ function depends only on one $\xi_X$, we keep using
$\Delta \chi^2 < 2.71$ and $\Delta \chi^2 < 6.63$ to determine the
90\%, 99\% CL contours, respectively.

As can be seen in fig. \ref{fig:scalar-vector-single-param}, among the
constraints implied by COHERENT data those for scalar-type
interactions are the most stringent. This can be understood as
follows. Although the data still involves large uncertainties, one can
see that it is rather consistent with SM expectations. The scalar
interaction involves a cross section that substantially differs from
that of the SM model, and so once added it worsens the
fit. Furthermore, translation from $\xi_S$ to quark parameters
involves nucleon mass fractions times nucleon-to-quark mass ratios
which altogether amount to values of order $5$ (see
eq. (\ref{eq:nuclear-coefficients})). Since $C_S$ is bounded from the
constraint derived on $\xi_S$, consistency demands a sort of
cancellation between the up and down couplings contributions, as
fig. \ref{fig:scalar-vector-single-param} (top panels) shows.  On the
other hand, tensor couplings allow for a relatively large freedom even
compared with vector parameters\footnote{This result differs from what
  has been found in ref. \cite{Kosmas:2017tsq}, the constraints
  derived here being less stringent.  The difference arises from the
  translation from nucleon to quark operators. In our case we use
  ``tensor charges'' (eq. (\ref{eq:nucleon-factors})), while
  ref. \cite{Kosmas:2017tsq} uses a vector-type translation.}, as
depicted in fig. \ref{fig:tensor-single-param}. This, however, does
not mean that tensor interactions provide a better fit to data than
vector do, as demonstrated by $\xi_T^\text{BFPV}=0$. It follows from
the translation from nucleus to quark parameters, which in the vector
case involve larger coefficients and so leads to narrower allowed
regions.  Finally, the presence of two minima in the
$\Delta \chi^2 (\xi_V)$ function translates into two separate linear
bands in the $C^u_V, C^d_V$ plane and in two concentric rings in the
$C^u_V, D^d_A$ plane.
%%%%%%%%%%%%%%%%%%%%%%%%%%%%%%%%%%%%%%%%%%%%%%
\subsection{two-parameter scenarios}
\label{sec:two}

\begin{table}
  \centering
  \renewcommand{\arraystretch}{1.4}
  \setlength{\tabcolsep}{0.9em}
  \begin{tabular}{|c|c|c|c|c|}\hline
    \textbf{Inter} & ($\xi_{X, 1},\xi_{X, 2})_\text{BFPV}$ & \textbf{P}
    &$90\%$ CL & $99\%$ CL\\\hline\hline
    \multirow{2}{*}{$\xi_S-\xi_V$} & $(-0.363,0.363)$ & $\xi_S$ & $[-0.816,0.816]$ & $[-1.123,1.123]$\\
                   &  $( -1.626, -0.253)$ & $\xi_V$ & $[-2.081,0.203]$ & $[-2.514,0.635]$\\\hline
    \multirow{2}{*}{$\xi_S-\xi_T$}& $0$ & $\xi_S$ &$[-0.623,0.623]$ & $[-1.076,1.076]$
    \\
    & $0$ & $\xi_T$ & $[-0.593,0.593]$ & $[-1.081,1.081]$ 
    \\\hline
    \multirow{2}{*}{$\xi_V-\xi_T$} &$(-1.398 ,-0.481)$ & $\xi_V$ & $[-2.081,0.203]$ & $[-2.51,0.632]$\\
     & $(-0.515, 0.515)$ & $\xi_T$ & $[-0.866,0.866]$ & $[-1.195,1.195]$
    \\\hline
  \end{tabular}
  \caption{Best-fit-point values for scalar, vector and tensor parameters 
    (second column), 90\% CL ($\Delta \chi^2 < 2.71$, fourth column) and 99\% CL ($\Delta \chi^2 < 6.63$, fifth column)
    ranges for $\xi_X$ as defined in eq. (\ref{eq:xi-definitions}). See text for more details.}
  \label{tab:xi-x-parameters-two-parameter-case}
\end{table}
\begin{figure}[!htb]
  \centering
  \includegraphics[scale=0.25]{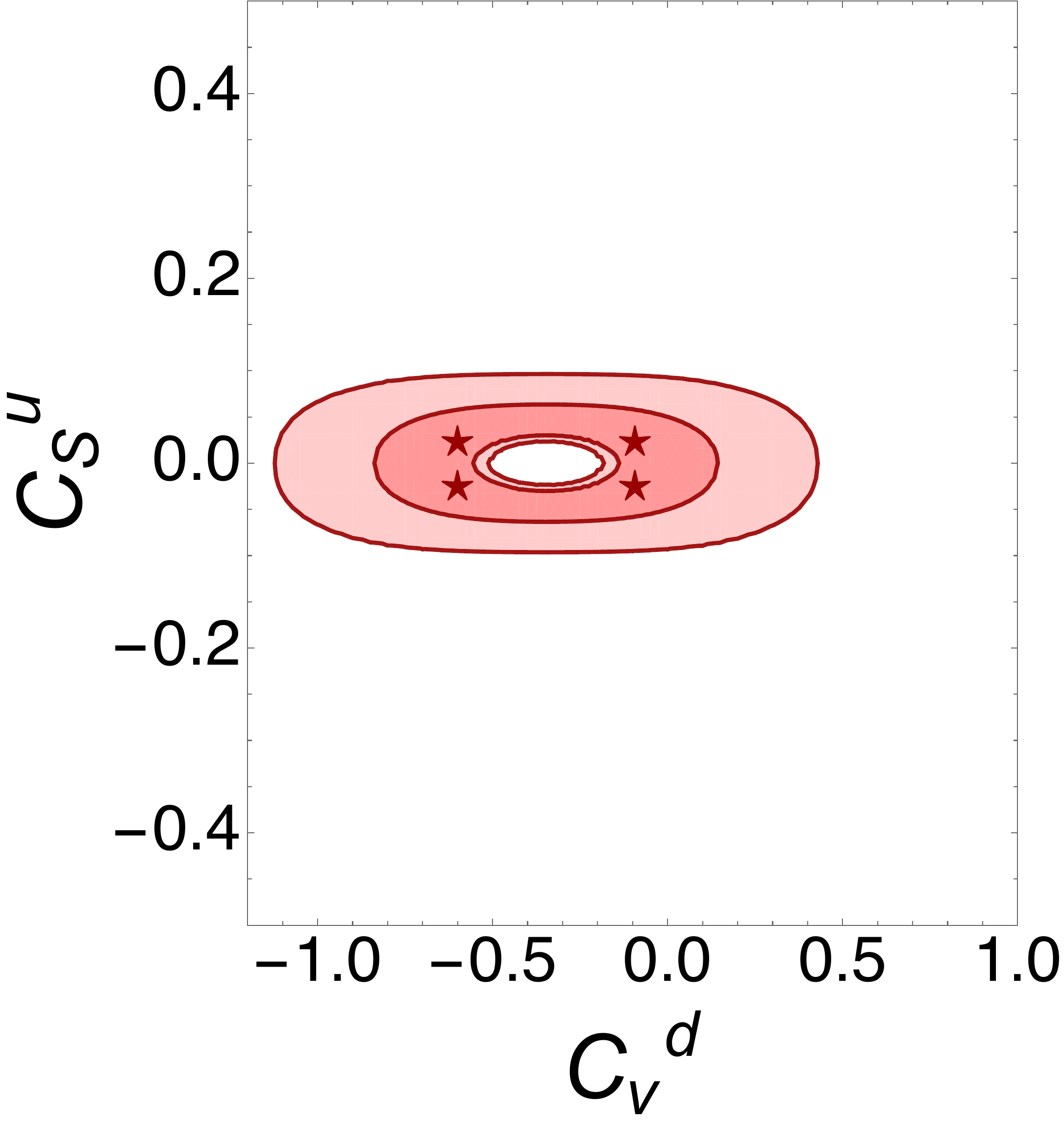}
  \includegraphics[scale=0.24]{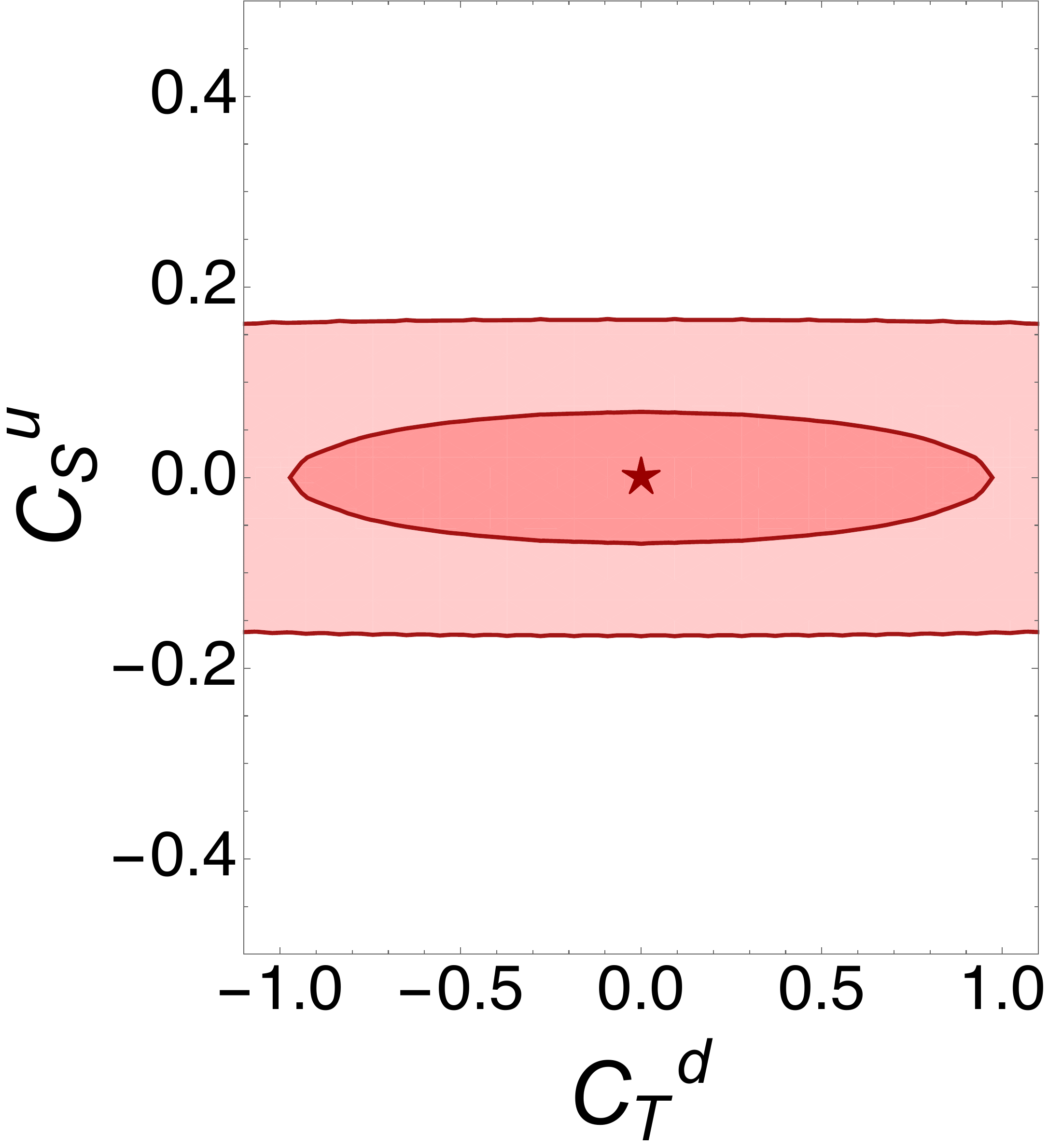}
  \includegraphics[scale=0.24]{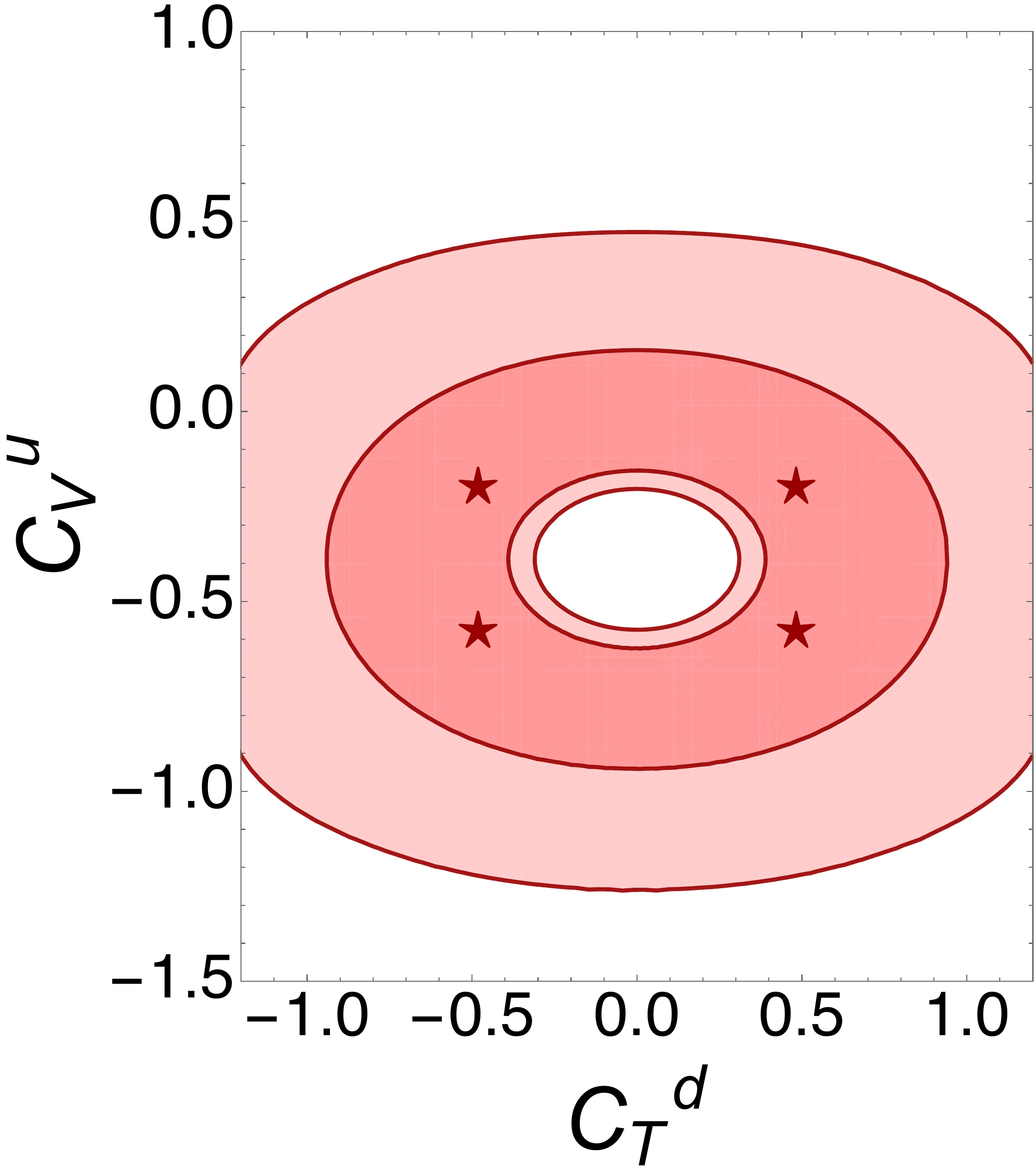}
  \caption{Results for the two parameter case analysis. Dark (light)
    reddish regions correspond to 90\%CL ($\Delta \chi^2 < 4.61$) and 99\%CL ($\Delta \chi^2 < 9.21$) bounds for the
    neutrino-quark couplings.}
  \label{fig:two-param}
\end{figure}
In this case we allow for the simultaneous presence of two
interactions at the nuclear level. Accordingly, we can distinguish
three cases corresponding to $\xi_S-\xi_V$, $\xi_S-\xi_T$ and
$\xi_V-\xi_T$ which involve eight and six quark parameters
respectively. As in the one-parameter case, here we focus on
smaller---though representative---regions of parameter space. To
determine at which extent the presence of a second interaction
modifies the constraints obtained in the single-parameter analysis we
study three scenarios: (2-i) for $\xi_S-\xi_V$, $C_S^u\neq 0$ and
$C_V^d \neq 0$; (2-ii) for $\xi_S-\xi_T$, $C_S^u\neq 0$
and $C_T^d \neq 0$; (2-iii) for $\xi_V-\xi_T$, $C_V^u\neq 0$ and $C_T^d \neq 0$.\\
The chi-square function for this analysis becomes now a function of
two parameters ($\xi_{X_1}$ and $\xi_{X_2}$).  We present in
tab.~\ref{tab:xi-x-parameters-two-parameter-case} the BFPVs and the
90\% and 99\% CL ranges for each $\xi_{X_i}$.  The CL ranges for the
parameter $\xi_{X_1}$ are obtained minimizing the least-squares
function over the nuisance parameters $\alpha$ and $\beta$ and over
the second interaction parameter $\xi_{X_2}$.  In principle, the
parameter $R$ could also be constrained by COHERENT data. However, its
contribution to the CE$\nu$NS cross section is subdominant with
respect to the SM contribution.  Moreover, it depends on the product
of two of the fundamental quark couplings $C^q_S,C^q_T$.  It turns out
that COHERENT bounds are not competitive enough to constrain $C^q_S$
and $C^q_T$ via the $R$ parameter, they are instead more stringently
constrained by the requirement of perturbativity (understood as
$C^q_{S,T}\leq 1$, i.e. the NGI should not exceed
$G_F$).  The constraints given in
tab.~\ref{tab:xi-x-parameters-two-parameter-case} can then be mapped
into the parameters of the neutrino-quark Lagrangian in the same way
as in the single-parameter analysis. Using the relations given in
eq. (\ref{eq:nuclear-coefficients}) we present in
fig. \ref{fig:two-param} the allowed regions for the fundamental
parameters in scenarios (2-i)-(2-iii).  We only show these three
particular cases, but the results in
tab. \ref{tab:xi-x-parameters-two-parameter-case} and
eq. (\ref{eq:nuclear-coefficients}) allow to investigate any case in
which two nuclear interactions are simultaneously present.

\begin{figure}[t]
  \centering
  \includegraphics[scale=0.45]{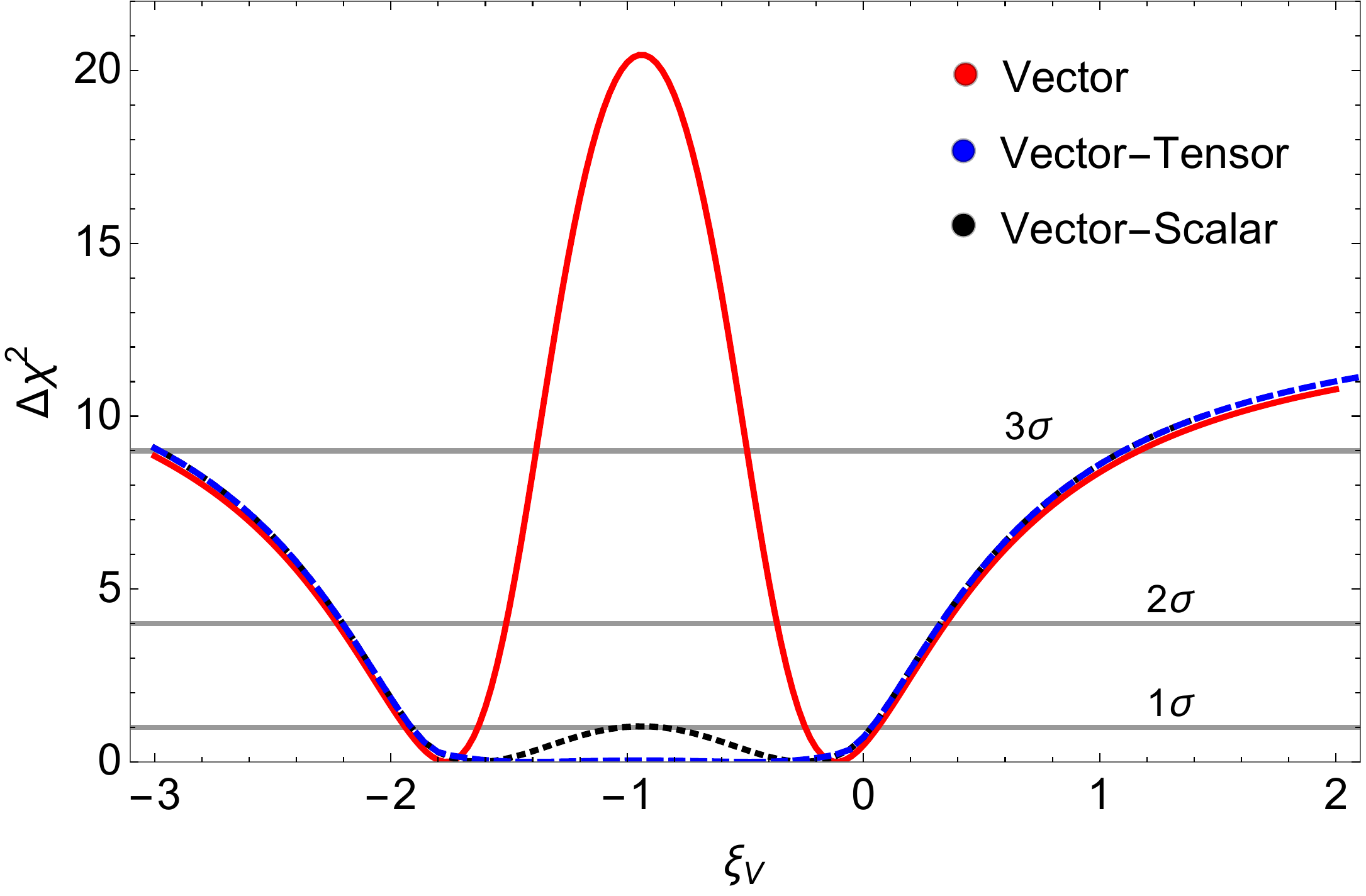}
  \caption{Dependence of the $\Delta \chi^2$ function on the $\xi_V$
    parameter in three different scenarios: The single-parameter
    scenario with only $\xi_V$ (red solid), the two-parameter
    scenario with $\xi_V$ and $\xi_T$ (blue dashed) and the
    two-parameter scenario with $\xi_V$ and $\xi_S$ (black dotted).}
  \label{fig:deltachi2_csiV}
\end{figure}

These results imply that the presence of an additional interaction at
the nuclear level relaxes the bounds on the fundamental neutrino-quark
couplings. Indeed, the addition of an extra free parameter $\xi_X$
allows for more freedom in the values of the NGI parameters.
Interestingly, COHERENT constraints on the vector interaction
parameter $\xi_V$ are sizeably relaxed with the addition of an extra
scalar or tensor interaction.  This can be seen by studying the
dependence of the $\Delta \chi^2$ function upon $\xi_V$, depicted in
fig.~\ref{fig:deltachi2_csiV}.  The red solid curve shows the
$\Delta \chi^2$ function in the single-parameter scenario where only
$\xi_V$ is switched on, while the blue dashed curve refers to the
two-parameter scenario with $\xi_V$ and $\xi_T$ simultaneously present
and the black dotted to the two-parameter scenario with $\xi_V$ and
$\xi_S$ both active.  In all three cases the $\Delta \chi^2$ function
has two minima, but the region between them is heavily modified when
an extra interaction is added.  In the region around
$\xi_V=-C_V=-[1-(1-4\sin^2\theta_w)-N/Z] \simeq -0.95$ the extra
vector interaction tends to cancel the SM contribution, thus worsening
the fit. As can be seen, the extra contribution (either scalar or
tensor) improves the fit by increasing the expected number of events
in that region.
\section{Conclusions}
\label{sec:concl}
We have studied a generic set of effective Lorentz invariant
non-derivative neutrino-quark interactions (NGI). These interactions
contain as a subset well-studied neutrino-quark NSI, but involve
additional scalar, pseudoscalar, axial and tensor couplings. In
contrast to vector interactions, they induce matter potentials that
are either helicity suppressed or vanish in non-polarized
media. Accordingly, they are poorly constrained by neutrino
oscillation data. They instead contribute to scattering processes
which set bounds on their values. We have considered the contributions
of NGI to the CE$\nu$NS process and we have employed the recent
COHERENT data to place constraints on the different effective
parameters.

Our analysis includes scalar, vector and tensor quark currents and
excludes pseudoscalar and axial quark couplings, which being
spin-dependent are expected to be less constrained. We have considered
diagonal as well as non-diagonal Lorentz structures, such as
$(\bar\nu\gamma_\mu\gamma_5\nu)(\bar q\gamma^\mu q)$ and
$(\bar\nu\gamma_5\nu)(\bar q\, q)$ and under the assumption of no
lepton flavor dependence and of a spin-1/2 nuclear ground state, we
have calculated the full CE$\nu$NS cross section for neutrinos and
anti-neutrinos. In order to assess the impact that such interactions
have on the CE$\nu$NS process, we have then carried out a chi-square
analysis in two simplified benchmark scenarios.  A first one where
only one nuclear interaction is present at a time, dubbed
single-parameter case, and a second where two are simultaneously
present, called two-parameter case.

In the single-parameter case, our findings show that the scalar
interaction is the most constrained, with the tightest bound found for
the Lorentz mixed pseudoscalar-scalar coupling. In such a case the
effective parameters are bounded to be smaller than $0.05$ at
90\%CL. For scalar-scalar couplings this bound is relaxed and the
parameters can be of order one, but still in a rather narrow region in
parameter space.  Allowed vector NGI are also sizable reaching values
as large as $0.85$ at 90\%CL, but again in two non-overlapping narrow
stripes. We find that tensor interactions are the less constrained,
with the reason being the translation between the nuclear to quark
parameters, which involves ``tensor charges'' which are small, thus
allowing for more freedom. Nevertheless none of these values lead to
an improvement in the COHERENT data fit, as the BFPVs found in our
analysis demonstrate.

In the two-parameter case, we have found that the presence of an
additional interaction at the nuclear level relaxes the bounds on the
fundamental neutrino-quark couplings.  The addition of an extra free
parameter $\xi_X$ allows the NGI to span over relatively larger
regions in parameter space. In particular, the allowed ranges for the
vector parameter $\xi_V$ are sizeably modified with the addition of an
extra scalar or tensor interaction. In the region where $\xi_V$ tends
to cancel the SM contribution, thus worsening the fit, the scalar or
tensor contribution enables its improvement to values below
$1\sigma$.

We have pointed out that further and perhaps more severe constraints
on NGI can be derived by considering instead DIS scattering data from
CHARM and NuTeV, as it turns out to be the case for neutrino NSI
\cite{Coloma:2017egw}. However, whether this is the case depends on
the mass of the mediator responsible for the effective interaction. We
have stressed that for mediator masses below $\sim 1\,$GeV our
constraints can be regarded as the current most stringent bounds on
NGI. For mediator masses above this value our results are still valid
but should be confronted with those from an analysis using DIS data,
which to our knowledge does not exist. At any rate, improvements on
limits on NGI couplings generated by mediators with masses below
$1\,$GeV will require further improvement of COHERENT data.

CE$\nu$NS offers a plethora of physics opportunities, allowing for
tests of anomalously large neutrino magnetic moments, sterile
neutrinos, new light degrees of freedom, among others
\cite{Billard:2018jnl}. The analysis presented in this paper, while
revisiting COHERENT constraints on some BSM interactions already
considered in the literature, further complements previous works by
considering effective NGI with mixed neutrino-quark Lorentz structures
and simultaneous presence of various neutrino-quark interactions, for
which we have shown that COHERENT data still allows for sizable
values.
\section*{Acknowledgments}
We thank Juan I. Collar and Bjorn Scholz for providing the COHERENT
data points and the final acceptance function. Grayson Rich for
providing the beam-on background data points as well as for clarifying
various aspects of their use.  We are also grateful to Juan Barranco,
Haris Kosmas, Mariam T\'ortola, Christoph A. Ternes, Luis Alvarez Ruso
and Juan M. Nieves for useful discussions.  DAS is supported by the
Chilean grant ``Unraveling new physics in the high-intensity and
high-energy frontiers'', Fondecyt No 1171136.  NR is funded by
proyecto FONDECYT Postdoctorado Nacional (2017) num. 3170135. VDR
acknowledges financial support by the ``Juan de la Cierva
Incorporaci\'on'' program (IJCI-2016-27736) funded by the Spanish
MINECO as well as partial support by the Spanish fellowship
Iberoam\'erica Santander Investigaci\'on 2017 and the Spanish grants
SEV-2014-0398 (MINECO), FPA2014-58183-P, PROMETEOII/2014/084
(Generalitat Valenciana).  VDR is also grateful for the kind
hospitality received at ``Universidad T\'ecnica Federico Santa Mar\'ia
(Campus Santiago San Joaqu\'in, Chile)'', during the initial stage of
this work.
\appendix
\section{Cross sections for neutrino and
  anti-neutrino CE$\nu$NS with spin-1/2 nuclei}
\label{sec:explicit-cal}
In this appendix we collect the results of the calculation of the
CE$\nu$NS cross section for neutrino and anti-neutrinos. We follow the
conventions used in ref. \cite{Lindner:2016wff}. We provide here the
full expression including all kinds of NGI, although in our analysis
we actually neglect pseudoscalar and axial nuclear currents.  We
compute the CE$\nu$NS cross section in the zero-momentum transfer
limit, that is when the spin-independent and spin-dependent nuclear
form factors satisfy $F^2(q\to 0)\to 1$ and $S^2(q\to 0)\to 1$.

\begin{figure}[t]
  \centering
  \includegraphics[scale=0.7]{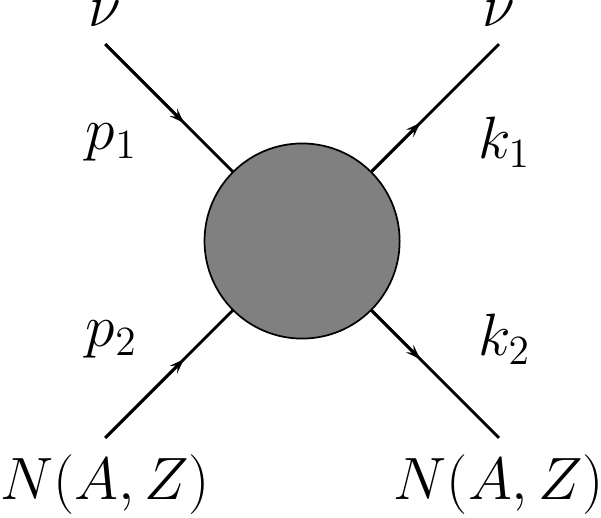}
  \caption{Diagram representing the CE$\nu$NS effective interaction
    momenta assignment.}
  \label{fig:cenues_diag}
\end{figure}

We begin the computation of the CE$\nu$NS cross section by describing
the kinematics of the process.  Incoming (outgoing)
neutrino/anti-neutrino and nucleus four-momenta are labeled $p_1$ and
$p_2$ ($k_1$ and $k_2$), as shown in fig.~\ref{fig:cenues_diag}. In
the lab-frame they are written as
\begin{alignat}{2}
  \label{eq:momenta}
  p_1&=(E_\nu,\hat e_z E_\nu)\ ,\qquad& k_1=(E_\nu^\prime,\hat e_r E_\nu^\prime)\ ,\\
  p_2&=(m_N,\vec{0})\ ,\qquad& k_2=(m_N,\hat e_r^\prime E_\nu^\prime)\ ,
\end{alignat}
where $\hat e_r=\hat e_z\cos\phi+\hat e_x\sin\phi$,
$\hat e_r^\prime=\hat e_z\cos\phi-\hat e_x\sin\phi$ and $\phi$ is the
scattering angle. The outgoing neutrino energy can be calculated as
\begin{equation}
  \label{eq:outgoing-energy}
  E_\nu^\prime=\frac{m_NE_\nu}{m_N+E_\nu(1-\cos\phi)}\ ,
\end{equation}
from which the nuclear recoil energy $E_r=E_\nu-E_\nu^\prime$ follows
\begin{equation}
  \label{eq:recoil-energy}
  E_r=\frac{E_\nu^2(1-\cos\phi)}{m_N+E_\nu(1-\cos\phi)}\ ,
\end{equation}
and from which in turn the maximum nuclear recoil energy is obtained
for backward scattering: $E_r^\text{max}\simeq 2E_\nu^2/m_N$
($E_\nu\ll m_N$).

The matrix elements for the process
$\bar\nu(p_1)+N(p_2)^{J=1/2}\to \bar\nu(k_1)+N(k_2)^{J=1/2}$ and
$\nu(p_1)+N(p_2)^{J=1/2}\to \nu(k_1)+N(k_2)^{J=1/2}$ can be written
according to
\begin{align}
  \label{eq:matrix-element-anti-neutrino}
  \mathcal{M}(\bar\nu + N \to \bar\nu + N)&=\frac{G_F}{\sqrt{2}}\sum_a
  \left[\bar v^s(p_1)P_R\Gamma^av^{s'}(k_1)\right]
  \left[\bar u^{r'}(k_2)\Gamma_a\left(C_a+i\gamma_5D_a\right)u^r(p_2)\right]\ ,
  \\
  \label{eq:matrix-element-anti-neutrino}
    \mathcal{M}(\nu + N \to \nu + N)&=\frac{G_F}{\sqrt{2}}\sum_a
  \left[\bar u^{s'}(k_1)\Gamma^aP_L u^{s}(p_1)\right]
  \left[\bar u^{r'}(k_2)\Gamma_a\left(C_a+i\gamma_5D_a\right)u^r(p_2)\right]\ .
\end{align}
Here $s,s',r,r'$ refer to spin indices and we sum over all Lorentz
structures. The differential cross section is in general given
by \cite{Patrignani:2016xqp}
\begin{equation}
  \label{eq:cross-section-generic}
  \frac{d\sigma}{dE_r}=\frac{1}{32\pi}\frac{1}{E_\nu^2m_N}
  \sum_{s,s'}\frac{1}{2}\sum_{r,r'}\left|\mathcal{M}^{s,s',r,r'}\right|^2\ ,
\end{equation}
where we have averaged over final state spins. Implementing the
kinematic relations in~(\ref{eq:momenta}) and using FeynCalc
\cite{Mertig:1990an,Shtabovenko:2016sxi} we arrive to the following
expressions (the result for the tensor interaction was derived, as far
as we know, for the first time in \cite{Barranco:2011wx})
\begin{align}
  \label{eq:cross-section-nu-anti-nu}
    \frac{d\sigma^a(q^2=0)}{dE_r}=
  \frac{G_F^2}{4\pi}m_{N_a}N_a^2
  &\left[\xi_S^2\,\frac{E_r }{E_r^\text{max}} 
  + |\vec\xi_V|^2\left(1 - \frac{E_r}{E_r^\text{max}} - 
  \frac{E_r}{E_\nu}\right)
  \pm 2\vec\xi_V\cdot\vec\xi_A\frac{E_r}{E_\nu}\right .
  \nonumber\\
  &+
  \left .
  |\vec\xi_A|^2
  \left(1 + \frac{E_r}{E_r^\text{max}}-\frac{E_r}{E_\nu}\right)
  +\xi_T^2
  \left(1 - \frac{E_r}{2 E_r^\text{max}}-\frac{E_r}{E_\nu}\right)
  \mp R\frac{E_r}{E_\nu} 
  \right]\ ,
\end{align}
where we have dropped $\mathcal{O}(E^2_r/E_\nu^2)$ terms. For
neutrinos the third and last terms are positive and negative
respectively, while for anti-neutrinos the signs are opposite. Here
the following conventions apply
\begin{alignat}{2}
  \label{eq:conv-x-sec}
  \xi_S^2&=\frac{C_S^2+D_S^2}{N^2}\ ,\qquad
  &\vec\xi_V&=\frac{C_V\hat e_V + D_A\hat e_A}{N}\ ,
  \\
  \vec\xi_A&=\frac{D_V\hat e_V + C_A\hat e_A}{N}\ ,\qquad
  &\xi_T^2&=8\frac{C_S^2+D_S^2}{N^2}
\end{alignat}
and
\begin{equation}
  \label{eq:R-param}
  R=2\frac{C_SC_T-C_PC_T+D_SD_T-D_PD_T}{N^2}\ .
\end{equation}
Our result for anti-neutrinos differs from that found in
ref. \cite{Lindner:2016wff} in the vector, axial and mixed
vector-axial terms (couplings). The energy dependence of those terms,
however, is the same and so the differences are numerically small.
\bibliography{references}

\providecommand{\href}[2]{#2}\begingroup\raggedright\begin{thebibliography}{10}

\bibitem{Akimov:2017ade}
{\scshape COHERENT} collaboration, D.~Akimov et~al., \emph{{Observation of
  Coherent Elastic Neutrino-Nucleus Scattering}},
  \href{http://dx.doi.org/10.1126/science.aao0990}{\emph{Science} (2017) },
  [\href{http://arxiv.org/abs/1708.01294}{{\tt 1708.01294}}].

\bibitem{Freedman:1973yd}
D.~Z. Freedman, \emph{{Coherent neutrino nucleus scattering as a probe of the
  weak neutral current}},
  \href{http://dx.doi.org/10.1103/PhysRevD.9.1389}{\emph{Phys. Rev.} {\bf D9}
  (1974) 1389--1392}.

\bibitem{Akimov:2015nza}
{\scshape COHERENT} collaboration, D.~Akimov et~al., \emph{{The COHERENT
  Experiment at the Spallation Neutron Source}},
  \href{http://arxiv.org/abs/1509.08702}{{\tt 1509.08702}}.

\bibitem{Freedman:1977xn}
D.~Z. Freedman, D.~N. Schramm and D.~L. Tubbs, \emph{{The Weak Neutral Current
  and Its Effects in Stellar Collapse}},
  \href{http://dx.doi.org/10.1146/annurev.ns.27.120177.001123}{\emph{Ann. Rev.
  Nucl. Part. Sci.} {\bf 27} (1977) 167--207}.

\bibitem{Brice:2013fwa}
S.~J. Brice et~al., \emph{{A method for measuring coherent elastic
  neutrino-nucleus scattering at a far off-axis high-energy neutrino beam
  target}}, \href{http://dx.doi.org/10.1103/PhysRevD.89.072004}{\emph{Phys.
  Rev.} {\bf D89} (2014) 072004}, [\href{http://arxiv.org/abs/1311.5958}{{\tt
  1311.5958}}].

\bibitem{Wong:2008vk}
H.~T. Wong, \emph{{Ultra-Low-Energy Germanium Detector for Neutrino-Nucleus
  Coherent Scattering and Dark Matter Searches}},
  \href{http://dx.doi.org/10.1142/S0217732308027801}{\emph{Mod. Phys. Lett.}
  {\bf A23} (2008) 1431--1442}, [\href{http://arxiv.org/abs/0803.0033}{{\tt
  0803.0033}}].

\bibitem{Aguilar-Arevalo:2016qen}
{\scshape CONNIE} collaboration, A.~Aguilar-Arevalo et~al., \emph{{Results of
  the engineering run of the Coherent Neutrino Nucleus Interaction Experiment
  (CONNIE)}},
  \href{http://dx.doi.org/10.1088/1748-0221/11/07/P07024}{\emph{JINST} {\bf 11}
  (2016) P07024}, [\href{http://arxiv.org/abs/1604.01343}{{\tt 1604.01343}}].

\bibitem{conus}
``The {CONUS} {E}xperiment.''
  \url{https://indico.cern.ch/event/606690/contributions/2591545/attachments/1499330/2336272/Taup2017_CONUS_talk_JHakenmueller.pdf}.

\bibitem{Kosmas:2017tsq}
D.~K. Papoulias and T.~S. Kosmas, \emph{{COHERENT constraints to conventional
  and exotic neutrino physics}},
  \href{http://dx.doi.org/10.1103/PhysRevD.97.033003}{\emph{Phys. Rev.} {\bf
  D97} (2018) 033003}, [\href{http://arxiv.org/abs/1711.09773}{{\tt
  1711.09773}}].

\bibitem{Canas:2018rng}
B.~C. Ca\~nas, E.~A. Garc\'es, O.~G. Miranda and A.~Parada, \emph{{Future
  perspectives for a weak mixing angle measurement in coherent elastic neutrino
  nucleus scattering experiments}},
  \href{http://arxiv.org/abs/1806.01310}{{\tt 1806.01310}}.

\bibitem{Vogel:1989iv}
P.~Vogel and J.~Engel, \emph{{Neutrino Electromagnetic Form-Factors}},
  \href{http://dx.doi.org/10.1103/PhysRevD.39.3378}{\emph{Phys. Rev.} {\bf D39}
  (1989) 3378}.

\bibitem{Scholberg:2005qs}
K.~Scholberg, \emph{{Prospects for measuring coherent neutrino-nucleus elastic
  scattering at a stopped-pion neutrino source}},
  \href{http://dx.doi.org/10.1103/PhysRevD.73.033005}{\emph{Phys. Rev.} {\bf
  D73} (2006) 033005}, [\href{http://arxiv.org/abs/hep-ex/0511042}{{\tt
  hep-ex/0511042}}].

\bibitem{Barranco:2005yy}
J.~Barranco, O.~G. Miranda and T.~I. Rashba, \emph{{Probing new physics with
  coherent neutrino scattering off nuclei}},
  \href{http://dx.doi.org/10.1088/1126-6708/2005/12/021}{\emph{JHEP} {\bf 12}
  (2005) 021}, [\href{http://arxiv.org/abs/hep-ph/0508299}{{\tt
  hep-ph/0508299}}].

\bibitem{Shoemaker:2017lzs}
I.~M. Shoemaker, \emph{{COHERENT search strategy for beyond standard model
  neutrino interactions}},
  \href{http://dx.doi.org/10.1103/PhysRevD.95.115028}{\emph{Phys. Rev.} {\bf
  D95} (2017) 115028}, [\href{http://arxiv.org/abs/1703.05774}{{\tt
  1703.05774}}].

\bibitem{Farzan:2018gtr}
Y.~Farzan, M.~Lindner, W.~Rodejohann and X.-J. Xu, \emph{{Probing neutrino
  coupling to a light scalar with coherent neutrino scattering}},
  \href{http://dx.doi.org/10.1007/JHEP05(2018)066}{\emph{JHEP} {\bf 05} (2018)
  066}, [\href{http://arxiv.org/abs/1802.05171}{{\tt 1802.05171}}].

\bibitem{Denton:2018xmq}
P.~B. Denton, Y.~Farzan and I.~M. Shoemaker, \emph{{A Plan to Rule out Large
  Non-Standard Neutrino Interactions After COHERENT Data}},
  \href{http://arxiv.org/abs/1804.03660}{{\tt 1804.03660}}.

\bibitem{Billard:2018jnl}
J.~Billard, J.~Johnston and B.~J. Kavanagh, \emph{{Prospects for exploring New
  Physics in Coherent Elastic Neutrino-Nucleus Scattering}},
  \href{http://arxiv.org/abs/1805.01798}{{\tt 1805.01798}}.

\bibitem{Horowitz:2003cz}
C.~J. Horowitz, K.~J. Coakley and D.~N. McKinsey, \emph{{Supernova observation
  via neutrino - nucleus elastic scattering in the CLEAN detector}},
  \href{http://dx.doi.org/10.1103/PhysRevD.68.023005}{\emph{Phys. Rev.} {\bf
  D68} (2003) 023005}, [\href{http://arxiv.org/abs/astro-ph/0302071}{{\tt
  astro-ph/0302071}}].

\bibitem{Patton:2012jr}
K.~Patton, J.~Engel, G.~C. McLaughlin and N.~Schunck, \emph{{Neutrino-nucleus
  coherent scattering as a probe of neutron density distributions}},
  \href{http://dx.doi.org/10.1103/PhysRevC.86.024612}{\emph{Phys. Rev.} {\bf
  C86} (2012) 024612}, [\href{http://arxiv.org/abs/1207.0693}{{\tt
  1207.0693}}].

\bibitem{Anderson:2012pn}
A.~J. Anderson, J.~M. Conrad, E.~Figueroa-Feliciano, C.~Ignarra, G.~Karagiorgi,
  K.~Scholberg et~al., \emph{{Measuring Active-to-Sterile Neutrino Oscillations
  with Neutral Current Coherent Neutrino-Nucleus Scattering}},
  \href{http://dx.doi.org/10.1103/PhysRevD.86.013004}{\emph{Phys. Rev.} {\bf
  D86} (2012) 013004}, [\href{http://arxiv.org/abs/1201.3805}{{\tt
  1201.3805}}].

\bibitem{Billard:2013qya}
J.~Billard, L.~Strigari and E.~Figueroa-Feliciano, \emph{{Implication of
  neutrino backgrounds on the reach of next generation dark matter direct
  detection experiments}},
  \href{http://dx.doi.org/10.1103/PhysRevD.89.023524}{\emph{Phys. Rev.} {\bf
  D89} (2014) 023524}, [\href{http://arxiv.org/abs/1307.5458}{{\tt
  1307.5458}}].

\bibitem{Essig:2018tss}
R.~Essig, M.~Sholapurkar and T.-T. Yu, \emph{{Solar Neutrinos as a Signal and
  Background in Direct-Detection Experiments Searching for Sub-GeV Dark Matter
  With Electron Recoils}},
  \href{http://dx.doi.org/10.1103/PhysRevD.97.095029}{\emph{Phys. Rev.} {\bf
  D97} (2018) 095029}, [\href{http://arxiv.org/abs/1801.10159}{{\tt
  1801.10159}}].

\bibitem{Dutta:2017nht}
B.~Dutta, S.~Liao, L.~E. Strigari and J.~W. Walker, \emph{{Non-standard
  interactions of solar neutrinos in dark matter experiments}},
  \href{http://dx.doi.org/10.1016/j.physletb.2017.08.031}{\emph{Phys. Lett.}
  {\bf B773} (2017) 242--246}, [\href{http://arxiv.org/abs/1705.00661}{{\tt
  1705.00661}}].

\bibitem{AristizabalSierra:2017joc}
D.~Aristizabal~Sierra, N.~Rojas and M.~H.~G. Tytgat, \emph{{Neutrino
  non-standard interactions and dark matter searches with multi-ton scale
  detectors}}, \href{http://dx.doi.org/10.1007/JHEP03(2018)197}{\emph{JHEP}
  {\bf 03} (2018) 197}, [\href{http://arxiv.org/abs/1712.09667}{{\tt
  1712.09667}}].

\bibitem{Gonzalez-Garcia:2018dep}
M.~C. Gonzalez-Garcia, M.~Maltoni, Y.~F. Perez-Gonzalez and
  R.~Zukanovich~Funchal, \emph{{Neutrino Discovery Limit of Dark Matter Direct
  Detection Experiments in the Presence of Non-Standard Interactions}},
  \href{http://arxiv.org/abs/1803.03650}{{\tt 1803.03650}}.

\bibitem{Papoulias:2018uzy}
D.~K. Papoulias, R.~Sahu, T.~S. Kosmas, V.~K.~B. Kota and B.~Nayak,
  \emph{{Novel neutrino-floor and dark matter searches with deformed shell
  model calculations}},  \href{http://arxiv.org/abs/1804.11319}{{\tt
  1804.11319}}.

\bibitem{Coloma:2017ncl}
P.~Coloma, M.~C. Gonzalez-Garcia, M.~Maltoni and T.~Schwetz, \emph{{A COHERENT
  enlightenment of the neutrino Dark Side}},
  \href{http://arxiv.org/abs/1708.02899}{{\tt 1708.02899}}.

\bibitem{Liao:2017uzy}
J.~Liao and D.~Marfatia, \emph{{COHERENT constraints on nonstandard neutrino
  interactions}},
  \href{http://dx.doi.org/10.1016/j.physletb.2017.10.046}{\emph{Phys. Lett.}
  {\bf B775} (2017) 54--57}, [\href{http://arxiv.org/abs/1708.04255}{{\tt
  1708.04255}}].

\bibitem{Cadeddu:2017etk}
M.~Cadeddu, C.~Giunti, Y.~F. Li and Y.~Y. Zhang, \emph{{Average CsI neutron
  density distribution from COHERENT data}},
  \href{http://dx.doi.org/10.1103/PhysRevLett.120.072501}{\emph{Phys. Rev.
  Lett.} {\bf 120} (2018) 072501}, [\href{http://arxiv.org/abs/1710.02730}{{\tt
  1710.02730}}].

\bibitem{Ge:2017mcq}
S.-F. Ge and I.~M. Shoemaker, \emph{{Constraining Photon Portal Dark Matter
  with Texono and Coherent Data}},  \href{http://arxiv.org/abs/1710.10889}{{\tt
  1710.10889}}.

\bibitem{Wolfenstein:1977ue}
L.~Wolfenstein, \emph{{Neutrino Oscillations in Matter}},
  \href{http://dx.doi.org/10.1103/PhysRevD.17.2369}{\emph{Phys. Rev.} {\bf D17}
  (1978) 2369--2374}.

\bibitem{Barger:2007im}
V.~Barger, P.~Langacker, M.~McCaskey, M.~J. Ramsey-Musolf and G.~Shaughnessy,
  \emph{{LHC Phenomenology of an Extended Standard Model with a Real Scalar
  Singlet}}, \href{http://dx.doi.org/10.1103/PhysRevD.77.035005}{\emph{Phys.
  Rev.} {\bf D77} (2008) 035005}, [\href{http://arxiv.org/abs/0706.4311}{{\tt
  0706.4311}}].

\bibitem{Lindner:2017uvt}
M.~Lindner, B.~Radovčić and J.~Welter, \emph{{Revisiting Large Neutrino
  Magnetic Moments}},
  \href{http://dx.doi.org/10.1007/JHEP07(2017)139}{\emph{JHEP} {\bf 07} (2017)
  139}, [\href{http://arxiv.org/abs/1706.02555}{{\tt 1706.02555}}].

\bibitem{Wise:2014oea}
M.~B. Wise and Y.~Zhang, \emph{{Effective Theory and Simple Completions for
  Neutrino Interactions}},
  \href{http://dx.doi.org/10.1103/PhysRevD.90.053005}{\emph{Phys. Rev.} {\bf
  D90} (2014) 053005}, [\href{http://arxiv.org/abs/1404.4663}{{\tt
  1404.4663}}].

\bibitem{Coloma:2017egw}
P.~Coloma, P.~B. Denton, M.~C. Gonzalez-Garcia, M.~Maltoni and T.~Schwetz,
  \emph{{Curtailing the Dark Side in Non-Standard Neutrino Interactions}},
  \href{http://dx.doi.org/10.1007/JHEP04(2017)116}{\emph{JHEP} {\bf 04} (2017)
  116}, [\href{http://arxiv.org/abs/1701.04828}{{\tt 1701.04828}}].

\bibitem{Akimov:2018vzs}
{\scshape COHERENT} collaboration, D.~Akimov et~al., \emph{{COHERENT
  Collaboration data release from the first observation of coherent elastic
  neutrino-nucleus scattering}},  \href{http://arxiv.org/abs/1804.09459}{{\tt
  1804.09459}}.

\bibitem{Lewin:1995rx}
J.~D. Lewin and P.~F. Smith, \emph{{Review of mathematics, numerical factors,
  and corrections for dark matter experiments based on elastic nuclear
  recoil}},
  \href{http://dx.doi.org/10.1016/S0927-6505(96)00047-3}{\emph{Astropart.
  Phys.} {\bf 6} (1996) 87--112}.

\bibitem{Escrihuela:2009up}
F.~J. Escrihuela, O.~G. Miranda, M.~A. Tortola and J.~W.~F. Valle,
  \emph{{Constraining nonstandard neutrino-quark interactions with solar,
  reactor and accelerator data}},
  \href{http://dx.doi.org/10.1103/PhysRevD.80.129908,
  10.1103/PhysRevD.80.105009}{\emph{Phys. Rev.} {\bf D80} (2009) 105009},
  [\href{http://arxiv.org/abs/0907.2630}{{\tt 0907.2630}}].

\bibitem{Bolanos:2008km}
A.~Bolanos, O.~G. Miranda, A.~Palazzo, M.~A. Tortola and J.~W.~F. Valle,
  \emph{{Probing non-standard neutrino-electron interactions with solar and
  reactor neutrinos}},
  \href{http://dx.doi.org/10.1103/PhysRevD.79.113012}{\emph{Phys. Rev.} {\bf
  D79} (2009) 113012}, [\href{http://arxiv.org/abs/0812.4417}{{\tt
  0812.4417}}].

\bibitem{Farzan:2017xzy}
Y.~Farzan and M.~Tortola, \emph{{Neutrino oscillations and Non-Standard
  Interactions}},
  \href{http://dx.doi.org/10.3389/fphy.2018.00010}{\emph{Front.in Phys.} {\bf
  6} (2018) 10}, [\href{http://arxiv.org/abs/1710.09360}{{\tt 1710.09360}}].

\bibitem{Lindner:2016wff}
M.~Lindner, W.~Rodejohann and X.-J. Xu, \emph{{Coherent Neutrino-Nucleus
  Scattering and new Neutrino Interactions}},
  \href{http://arxiv.org/abs/1612.04150}{{\tt 1612.04150}}.

\bibitem{Fitzpatrick:2012ix}
A.~L. Fitzpatrick, W.~Haxton, E.~Katz, N.~Lubbers and Y.~Xu, \emph{{The
  Effective Field Theory of Dark Matter Direct Detection}},
  \href{http://dx.doi.org/10.1088/1475-7516/2013/02/004}{\emph{JCAP} {\bf 1302}
  (2013) 004}, [\href{http://arxiv.org/abs/1203.3542}{{\tt 1203.3542}}].

\bibitem{Dent:2015zpa}
J.~B. Dent, L.~M. Krauss, J.~L. Newstead and S.~Sabharwal, \emph{{General
  analysis of direct dark matter detection: From microphysics to observational
  signatures}}, \href{http://dx.doi.org/10.1103/PhysRevD.92.063515}{\emph{Phys.
  Rev.} {\bf D92} (2015) 063515}, [\href{http://arxiv.org/abs/1505.03117}{{\tt
  1505.03117}}].

\bibitem{DelNobile:2013sia}
M.~Cirelli, E.~Del~Nobile and P.~Panci, \emph{{Tools for model-independent
  bounds in direct dark matter searches}},
  \href{http://dx.doi.org/10.1088/1475-7516/2013/10/019}{\emph{JCAP} {\bf 1310}
  (2013) 019}, [\href{http://arxiv.org/abs/1307.5955}{{\tt 1307.5955}}].

\bibitem{Cheng:1988im}
H.-Y. Cheng, \emph{{Low-energy Interactions of Scalar and Pseudoscalar Higgs
  Bosons With Baryons}},
  \href{http://dx.doi.org/10.1016/0370-2693(89)90402-4}{\emph{Phys. Lett.} {\bf
  B219} (1989) 347--353}.

\bibitem{Anselmino:2008jk}
M.~Anselmino, M.~Boglione, U.~D'Alesio, A.~Kotzinian, F.~Murgia, A.~Prokudin
  et~al., \emph{{Update on transversity and Collins functions from SIDIS and e+
  e- data}},
  \href{http://dx.doi.org/10.1016/j.nuclphysbps.2009.03.117}{\emph{Nucl. Phys.
  Proc. Suppl.} {\bf 191} (2009) 98--107},
  [\href{http://arxiv.org/abs/0812.4366}{{\tt 0812.4366}}].

\bibitem{Courtoy:2015haa}
A.~Courtoy, S.~Baeßler, M.~González-Alonso and S.~Liuti,
  \emph{{Beyond-Standard-Model Tensor Interaction and Hadron Phenomenology}},
  \href{http://dx.doi.org/10.1103/PhysRevLett.115.162001}{\emph{Phys. Rev.
  Lett.} {\bf 115} (2015) 162001}, [\href{http://arxiv.org/abs/1503.06814}{{\tt
  1503.06814}}].

\bibitem{Goldstein:2014aja}
G.~R. Goldstein, J.~O. Gonzalez~Hernandez and S.~Liuti, \emph{{Flavor
  dependence of chiral odd generalized parton distributions and the tensor
  charge from the analysis of combined $\pi^0$ and $\eta$ exclusive
  electroproduction data}},  \href{http://arxiv.org/abs/1401.0438}{{\tt
  1401.0438}}.

\bibitem{Radici:2015mwa}
M.~Radici, A.~Courtoy, A.~Bacchetta and M.~Guagnelli, \emph{{Improved
  extraction of valence transversity distributions from inclusive dihadron
  production}}, \href{http://dx.doi.org/10.1007/JHEP05(2015)123}{\emph{JHEP}
  {\bf 05} (2015) 123}, [\href{http://arxiv.org/abs/1503.03495}{{\tt
  1503.03495}}].

\bibitem{Jungman:1995df}
G.~Jungman, M.~Kamionkowski and K.~Griest, \emph{{Supersymmetric dark matter}},
  \href{http://dx.doi.org/10.1016/0370-1573(95)00058-5}{\emph{Phys. Rept.} {\bf
  267} (1996) 195--373}, [\href{http://arxiv.org/abs/hep-ph/9506380}{{\tt
  hep-ph/9506380}}].

\bibitem{Dorenbosch:1986tb}
{\scshape CHARM} collaboration, J.~Dorenbosch et~al., \emph{{Experimental
  Verification of the Universality of $\nu_e$ and $\nu_\mu$ Coupling to the
  Neutral Weak Current}},
  \href{http://dx.doi.org/10.1016/0370-2693(86)90315-1}{\emph{Phys. Lett.} {\bf
  B180} (1986) 303--307}.

\bibitem{Zeller:2001hh}
{\scshape NuTeV} collaboration, G.~P. Zeller et~al., \emph{{A Precise
  determination of electroweak parameters in neutrino nucleon scattering}},
  \href{http://dx.doi.org/10.1103/PhysRevLett.88.091802}{\emph{Phys. Rev.
  Lett.} {\bf 88} (2002) 091802},
  [\href{http://arxiv.org/abs/hep-ex/0110059}{{\tt hep-ex/0110059}}].

\bibitem{Mikheev:1986gs}
S.~P. Mikheev and A.~{\relax Yu}. Smirnov, \emph{{Resonance Amplification of
  Oscillations in Matter and Spectroscopy of Solar Neutrinos}}, {\emph{Sov. J.
  Nucl. Phys.} {\bf 42} (1985) 913--917}.

\bibitem{Mikheev:1986wj}
S.~P. Mikheev and A.~{\relax Yu}. Smirnov, \emph{{Resonant amplification of
  neutrino oscillations in matter and solar neutrino spectroscopy}},
  \href{http://dx.doi.org/10.1007/BF02508049}{\emph{Nuovo Cim.} {\bf C9} (1986)
  17--26}.

\bibitem{Esteban:2018ppq}
I.~Esteban, M.~C. Gonzalez-Garcia, M.~Maltoni, I.~Martinez-Soler and
  J.~Salvado, \emph{{Updated Constraints on Non-Standard Interactions from
  Global Analysis of Oscillation Data}},
  \href{http://arxiv.org/abs/1805.04530}{{\tt 1805.04530}}.

\bibitem{Bergmann:1999rz}
S.~Bergmann, Y.~Grossman and E.~Nardi, \emph{{Neutrino propagation in matter
  with general interactions}},
  \href{http://dx.doi.org/10.1103/PhysRevD.60.093008}{\emph{Phys. Rev.} {\bf
  D60} (1999) 093008}, [\href{http://arxiv.org/abs/hep-ph/9903517}{{\tt
  hep-ph/9903517}}].

\bibitem{Patrignani:2016xqp}
{\scshape Particle Data Group} collaboration, C.~Patrignani et~al.,
  \emph{{Review of Particle Physics}},
  \href{http://dx.doi.org/10.1088/1674-1137/40/10/100001}{\emph{Chin. Phys.}
  {\bf C40} (2016) 100001}.

\bibitem{Mertig:1990an}
R.~Mertig, M.~Bohm and A.~Denner, \emph{{FEYN CALC: Computer algebraic
  calculation of Feynman amplitudes}},
  \href{http://dx.doi.org/10.1016/0010-4655(91)90130-D}{\emph{Comput. Phys.
  Commun.} {\bf 64} (1991) 345--359}.

\bibitem{Shtabovenko:2016sxi}
V.~Shtabovenko, R.~Mertig and F.~Orellana, \emph{{New Developments in FeynCalc
  9.0}}, \href{http://dx.doi.org/10.1016/j.cpc.2016.06.008}{\emph{Comput. Phys.
  Commun.} {\bf 207} (2016) 432--444},
  [\href{http://arxiv.org/abs/1601.01167}{{\tt 1601.01167}}].

\bibitem{Barranco:2011wx}
J.~Barranco, A.~Bolanos, E.~A. Garces, O.~G. Miranda and T.~I. Rashba,
  \emph{{Tensorial NSI and Unparticle physics in neutrino scattering}},
  \href{http://dx.doi.org/10.1142/S0217751X12501473}{\emph{Int. J. Mod. Phys.}
  {\bf A27} (2012) 1250147}, [\href{http://arxiv.org/abs/1108.1220}{{\tt
  1108.1220}}].

\end{thebibliography}\endgroup
\end{document}